\documentclass[useAMS,usenatbib]{mn2e}
\usepackage{graphicx}
\usepackage{threeparttable}
\usepackage{tablefootnote}
\usepackage{amsmath}
\usepackage{color}
\usepackage{float}
\usepackage{caption}
\usepackage{placeins}
\usepackage{multicol}
\usepackage{soul}

\def\be{\begin{equation}}
\def\ee{\end{equation}}

\def\msun{{\Msun}}

\def\HI{\hbox{H~$\scriptstyle\rm I\ $}}

\def\gsim{\lower.5ex\hbox{\gtsima}}
\def\lsim{\lower.5ex\hbox{\ltsima}} \def\gtsima{$\; \buildrel > \over
\sim \;$} \def\ltsima{$\; \buildrel < \over \sim \;$} \def\prosima{$\;
\buildrel \propto \over \sim \;$} \def\gsim{\lower.5ex\hbox{\gtsima}}
\def\lsim{\lower.5ex\hbox{\ltsima}}
\def\simgt{\lower.5ex\hbox{\gtsima}}
\def\simlt{\lower.5ex\hbox{\ltsima}}
\def\simpr{\lower.5ex\hbox{\prosima}} \def\la{\lsim} \def\ga{\gsim}
 
 \def\gtsima{$\; \buildrel > \over \sim \;$}
\def\ltsima{$\; \buildrel < \over \sim \;$}
\def\gsim{\lower.5ex\hbox{\gtsima}}
\def\lsim{\lower.5ex\hbox{\ltsima}}
\def\simgt{\lower.5ex\hbox{\gtsima}}
\def\simlt{\lower.5ex\hbox{\ltsima}}
\def\simpr{\lower.5ex\hbox{\prosima}}
\def\la{\lsim}
\def\ga{\gsim}

\def\msun{\,{\rm \Msun}}

\def\E3{{\cal E}_{\rm g}^{III}}

\def\Msun{\rm M_\odot}

\def\Zsun{\rm Z_\odot}

\def\Msun{\rm M_\odot}

\def\Zsun{\rm Z_\odot}
\def\M*{M_*}

\def\Z*{Z_*}
\def\L*{L_*}
\def\muv{\rm M_{UV}}

\def\fesc{f_{esc}}

\def\feff{f_*^{\rm eff}}

\newcommand{\quotes}[1]{``#1''}
\newcommand\textlcsc[1]{\textsc{\MakeLowercase{#1}}}

\title[Cosmic variance during the EoR]{Astraeus II: Quantifying the impact of cosmic variance during the Epoch of Reionization}
\author[G. Ucci et al.]{Graziano Ucci$^1$\thanks{g.ucci@rug.nl}, Pratika Dayal$^1$, Anne Hutter$^1$, Gustavo Yepes$^{2,3}$
\newauthor Stefan Gottl{\"o}ber$^4$, Laurent Legrand$^1$, Laura Pentericci$^5$
\newauthor Marco Castellano$^5$, Tirthankar Roy Choudhury$^6$\\
$^{1}$ Kapteyn Astronomical Institute, University of Groningen, P.O. Box 800, 9700 AV Groningen, The Netherlands \\
$^{2}$ Departamento de Fısica Teorica, Modulo 8, Facultad de Ciencias, Universidad Autonoma de Madrid, 28049 Madrid, Spain\\
$^{3}$ CIAFF, Facultad de Ciencias, Universidad Autonoma de Madrid, 28049 Madrid, Spain\\
$^{4}$ Leibniz-Institut für Astrophysik, An der Sternwarte 16, 14482 Potsdam, Germany\\
$^{5}$ INAF-OAR, Via Frascati 33, Monte Porzio Catone (RM), Italy \\
$^{6}$ National Centre for Radio Astrophysics, Tata Institute of Fundamental Research, Pune 411007, India}
\begin{document}

\date{}

\maketitle

\begin{abstract}
Next generation telescopes such as the James Webb Space Telescope (JWST) and the Nancy Grace Roman Space Telescope (NGRST) will enable us to study the first billion years of our Universe in unprecedented detail. In this work we use the \textlcsc{Astraeus} (semi-numerical rAdiative tranSfer coupling of galaxy formaTion and Reionization in N-body dArk mattEr simUlationS) framework, that couples galaxy formation and reionization (for a wide range of reionization feedback models), to estimate the cosmic variance expected in the UV Luminosity Function (UV LF) and the Stellar Mass Function (SMF) in JWST surveys. We find that different reionization scenarios play a minor role in the cosmic variance. Most of the cosmic variance is completely driven by the underlying density field and increases above $100\%$ for $\muv \sim -17.5 ~ (-20)$ at $z =12~ (6)$ for the JADES-deep survey (the deep JWST Advanced Extragalactic Survey with an area of 46 arcmin$^2$); the cosmic variance decreases with an increasing survey area roughly independently of redshift. We find that the faint-end ($\muv \gsim -17$) slope of the Lyman Break Galaxies (LBGs) UV LF becomes increasingly shallower with increasing reionization feedback and show how JWST observations will be able to distinguish between different models of reionization feedback at $z>9$, even accounting for cosmic variance. We also show the environments (in terms of density and ionization fields)of Lyman Break Galaxies during the EoR, finding that the underlying over-density and ionization fraction scale positively with the UV luminosity. Finally, we also provide a public software tool to allow interested readers to compute cosmic variance for different redshifts and survey areas.
\end{abstract}


\section{Introduction}
One of the most important questions in physical cosmology regards the ionization of neutral hydrogen (\HI) in the intergalactic medium (IGM) within the first billion years of the Universe. During this era, termed the \quotes{Epoch of Reionization} (EoR), Lyman continuum photons from early galaxy populations started ionizing the \HI in their vicinity. This led to large fluctuations in the spatial and temporal distribution of ionized regions, resulting in reionization being a highly patchy process. A growing body of theoretical \citep[e.g.][]{choudhury2007,razoumov2010,salvaterra2011,liu2016, qin2017, dayal2017b, astraeus1} and observational works \citep{finkelstein2012, bouwens2012, duncan2015, robertson2015} converge on faint galaxies (with absolute magnitudes $\muv \gsim -15$) being the key reionization sources. Additionally, number of works also find that Active Galactic Nuclei (AGN) could have had a non-negligible contribution to reionization \citep[e.g.,][]{volonteri2009, giallongo2015,madau2015, chardin2017, mitra2018, seiler2018, finkelstein2019, dayal2020}. The complex topology and sources of reionization, therefore, still remain compelling open questions. A key reason for this is that the redshift-dependent reionization contribution from star formation and AGN crucially depends on a number of (poorly known) parameters. These include the minimum halo mass of star-forming and AGN hosting galaxies, the intrinsic production rate of \HI ionizing photons from star formation and AGN, the escape fraction ($f_{\rm esc}$) of \HI ionizing photons from the galactic environment (into the IGM), the impact of the heating ultra-violet background (UVB) created during reionization on the gas content of low-mass halos and the IGM clumping factor during the EoR \citep[for details see Sec. 7 in][]{dayal2018}.

The past decade has seen an enormous increase in the data collected for high-redshift Lyman break Galaxies (LBGs) in the EoR using both space- and ground-based facilities, such as the William Herschel Telescope (WHT)\footnote{\scriptsize https://www.esa.int/Science\_Exploration/Space\_Science/Herschel\_overview}, the Hubble Space Telescope (HST)\footnote{\scriptsize https://www.nasa.gov/mission\_pages/hubble/main/index.html} and the Very Large Telescope (VLT)\footnote{\scriptsize https://www.eso.org/public/unitedkingdom/teles-instr/paranal-observatory/vlt/?lang}, to name a few. This has resulted in statistically significant ultraviolet luminosity function (UV LF) extending to UV magnitudes as faint as $\muv \sim -17$ for blank fields \citep{mclure2013,bouwens2015b} and $\muv \sim -14$ for lensed fields \citep[e.g.,][]{livermore2017,atek2018,ishigaki2018,castellano2016,yue2018} at $z \sim 7$, although the latter are susceptible to relatively large uncertainties due to lens modelling and completeness corrections. Additionally, these observations have allowed estimates of the stellar mass function (SMF) extending down to stellar masses of $M_* \sim 10^7\msun$ at $z\sim 6-8$ \citep{gonzalez2011,duncan2014,song2016}.

In the near future, next generation facilities, such as the James Webb Space Telescope (JWST)\footnote{\scriptsize https://www.jwst.nasa.gov/} and the Nancy Grace Roman Space Telescope (NGRST, formerly known as WFIRST)\footnote{\scriptsize https://www.nasa.gov/content/goddard/about-nancy-grace-roman-space-telescope}, will enable us to study galaxies in the first billion years of the Universe in unprecedented detail. Indeed, JWST will enable resolved spectroscopy of early galaxies \citep[although limited to bright and rare objects; e.g.,][]{stark2016,williams2018}, and extend the UV LF by pushing up to two magnitudes deeper than current observations with the HST \citep{finkelstein2015arXiv,williams2018,kemp2019,rieke2019}.

However, as a result of the small-scale inhomogeneities in the density distribution, measurements of galaxy properties (such as the UV LF and SMF) in small-volume observations are greatly affected by cosmic variance. This is a dominant source of error in many extragalactic measurements due to the limited volumes/luminosities probed. Furthermore, during the EoR,  the impact of reionization feedback on galaxy formation varies as a function of spatial position and time \citep[e.g.,][]{dawoodbhoy2018}. Indeed, in ionized regions, the faint-end slope of the UV LF becomes shallower as a result of the lower-gas masses left in low-mass halos due to the reionization background  \citep[e.g.,][]{choudhury2019}. Quantifying the cosmic variance for forthcoming galaxy surveys in the EoR is therefore crucial before they can be used to study the physics of the underlying galaxy population.

A number of works have presented cosmic variance results at high-$z$: using analytic estimates via the two-point correlation function in extended Press-Schechter theory and mock catalogues obtained from N-body cosmological simulations (that use a mass-luminosity relation to link galaxies to the underlying halo masses), \citet{Trenti2008} studied the impact of cosmic variance on the observed UV LF. \citet{Moster2011} have studied the cosmic variance in the Dark Matter distribution for different survey geometries as a function of mean redshift and redshift bin size by integrating the correlation function for a pencil beam geometry. On the other hand, \citet{Bhowmick2019} used \textlcsc{BlueTides}, a large and high resolution cosmological hydrodynamical simulation run until $z \sim 7.5$ to directly estimate the cosmic variance for current and upcoming surveys. Taking advantage of the \textlcsc{BlueTides} volume and resolution (400$^3$ Mpc/h$^3$ and $2 \times 7048^3$ particles), they probed the bias and cosmic variance of $z > 7$ galaxies between magnitude H $\sim 30-25$ over survey areas of $\sim$ 0.1 arcmin$^2$ to 10 deg$^2$. \textlcsc{BlueTides} includes various sub-grid physics models (multiphase star formation model, molecular hydrogen formation, gas and metal cooling, supernovae feedback, black hole feedback growth and AGN feedback) in addition to an analytic model for patchy reionization. While our work is similar in spirit to that of \citet{Bhowmick2019}, in addition to using a semi-numerical radiative feedback model for reionization, we couple reionization feedback to galaxy formation at high redshift.

In this work, we use the \textlcsc{Astraeus} framework to quantify the cosmic variance in the halo mass function (HMF), UV LF and SMF for forthcoming surveys with the JWST and NGRST. This model couples a state-of-the-art N-body simulation with a semi-analytic model of galaxy formation and a semi-numerical radiative transfer code for reionization \citep[cf. also models in][]{mutch2016,seiler2018}. The key strengths of the model lie in ({\it i}) the large volume that is modelled ($160 h^{-1}$ comoving Mpc), yielding a statistically significant number of JWST fields; ({\it ii}) the wide range of UV magnitudes explored ($-22 < \muv < -8$)\footnote{We caution the reader that the results for $\muv \gsim -12$ are somewhat limited by halo mass resolution and are therefore not complete in a statistical sense.}; and ({\it iii}) the large range of reionization feedback scenarios that are studied that delimit the physically plausible range for the UV LFs and SMFs. In addition, our framework also yield hints on the redshift evolution of the environment of LBGs, specially in terms of the underlying density and ionization fields \citep[see also][]{hutter2017}.

Throughout this work we assumed a Salpeter initial mass function \citep[IMF,][]{salpeter1955} between 0.1 and 100 $\Msun$, and the following set of cosmological parameters: $\Omega_\Lambda = 0.69$, $\Omega_M = 0.31$, $h = 0.6777$, $n_s = 0.96$, $\sigma_8 = 0.82$ \citep{planck2016}.

\begin{table*}
	\caption{For the UV feedback models noted in column 1, we show the value of the threshold star formation efficiency (column 2), the fraction of SNII energy that can couple to gas (column 3), the escape fraction of \HI ionizing photons (column 4), the IGM temperature in ionized regions (column 5) and the characteristic halo mass $M_c$ at $z = 7$ for which halos can retain half of their gas mass after UV feedback assuming a reionization redshift $z_{reion} = 8~ (12)$ (column 6).}
	\begin{threeparttable}
	\begin{tabular}{c c c c c c}
		\hline\hline
		Model & ${f_*}$ & $f_w$ & $f_\mathrm{esc}$ & $T_{\rm IGM}$ & log ($M_c / \msun$) \\
		\hline
		\textit{Early heating} & 0.01 & 0.2 & $0.6 \times \min\left[1, f_\star \left(1 + \frac{f_w E_{51} \nu_z}{\left(3\pi G H_0\right)^{2/3} \Omega_m^{1/3} (1+z) M_h^{2/3}}\right) \right]$\footnotemark[1] & $4 \times 10^4$~K & 9.08\footnotemark[2] (7.60)\\
		\textit{Photo-ionization} & 0.01 & 0.2 & 0.215 & $\sim 4 \times 10^4$~K\footnotemark[4] & 9.09\footnotemark[3] (7.86)\\
		\textit{Jeans mass} & 0.01 & 0.2 & 0.285 & $4 \times 10^4$~K & 9.52 (9.52)\\
		\hline
		\end{tabular}
\begin{tablenotes}
\item[1]{\scriptsize \citep{astraeus1}; instantaneous SN feedback}
\item[2]{\scriptsize \citep{gnedin2000,naoz2013}}
\item[3]{\scriptsize \citep{sobacchi2013a}}
\item[4]{\scriptsize $T_\mathrm{IGM}$ is actually given by the \HI photoionization rate $\Gamma_\mathrm{HI}$ which is about $\sim10^{-12.3}$s$^{-1}$.}
\end{tablenotes}
\end{threeparttable}
\label{table:models}
\end{table*}

\section{Theoretical Model}\label{sec:model}
In this work we use the \textlcsc{Astraeus} (semi-numerical rAdiative tranSfer coupling of galaxy formaTion and Reionization in N-body dArk mattEr simUlationS) framework that couples a state-of-the-art N-body simulation run as part of the Multi-dark project\footnote{See www.cosmosim.org for further information about the Multi-dark suite of simulations and access to the simulations database.} ({\it Very small multi-dark Planck}; \textlcsc{VSMDPL}) with a slightly modified version of the \textlcsc{Delphi} semi-analytic model of galaxy formation \citep{dayal2014,dayal2015,dayal2017a} and the \textlcsc{CIFOG} (Code to compute ionization field from density fields and source catalogue) semi-numerical reionization scheme \citep{hutter2018}. We briefly discuss the framework here and interested readers are referred to \citep{astraeus1} for complete details.

The N-body simulation was run with a box size of 160$h^{-1}$ Mpc and 3840$^3$ dark matter particles resulting in a resolution mass of $6.2 \times 10^6 h^{-1}$ $\Msun$. The equivalent Plummer's gravitational softening was set to $2 h^{-1}$ physical kpc. A total of 150 different snapshots of the simulation were stored between $z=25$ and $z=0$. The \textlcsc{Rockstar}  phase-space  halo finder \citep{behroozi2013} was used to identify all halos and subhalos in each snapshot, down to 20 particles per halo, resulting in a minimum resolved halo mass of $M_h = 1.24 \times 10^8 h^{-1}$ $\msun$. In addition, merger trees from the \textlcsc{Rockstar} halo catalogues were computed using the \textlcsc{Consistent trees} \citep{behroozi2013_trees} method. The outputs were re-sorted locally horizontally (i.e. on a tree by tree basis but within in a tree by redshift) allowing the flexibility of using \textlcsc{Astraeus} as fully vertical or horizontal semi-analytic galaxy formation framework.

Our semi-analytic galaxy formation model \textlcsc{Delphi} includes all the key baryonic processes of gas accretion, gas and stellar mass being brought in by mergers, star formation and the associated Type II supernova (SNII) feedback as detailed below. We also include the impact of radiative feedback from the patchy UVB generated during reionization. The UVB can photo-evaporate gas from small gravitational potentials and increase the Jeans mass for galaxy formation (the latter reduces the amount of gas accreted from the IGM), leading to a reduction in the gas content of low-mass halos in ionized regions. At each redshift-step, these baryonic processes are coupled to the merger- and accretion-driven growth of the dark matter halos obtained from the N-body simulations as detailed in what follows. Our model has a total of three free parameters: the threshold star formation efficiency ($f_\star$), the fraction of SNII energy that couples to gas ($f_w$) and the escape fraction of ionizing photons from the galactic environment into the IGM ($f_\mathrm{esc}$). These  are tuned to reproduce the key observables for both galaxies (the UV LFs and SMFs at $z = 5 - 10$, the star formation rate density evolution) and reionization (the electron scattering optical depth and the ionizing history constraints from quasars, Lyman Alpha Emitters and Gamma Ray Bursts). The values of the model free parameters, the IGM temperature in ionized regions and the characteristic halo mass for each of the UV feedback models used in this paper are reported in Table \ref{table:models}.

\begin{itemize}
\item {\it Initial gas mass}: halos that have no progenitors are assigned a cosmological ratio ($\Omega_b/\Omega_m$) of gas-to-dark matter mass smoothly accreted from the IGM. Halos that have progenitors gain gas both through mergers and smooth accretion from the IGM. For halos in ionized regions, this gas mass can be reduced (depending on the halo mass, its redshift, the UVB strength and the redshift at which the halo was first irradiated by the UVB) due to reionization feedback as detailed in what follows.\\

\item {\it Star formation}: at a given redshift step the initial gas mass can form stars with an {\it effective} efficiency $\feff$ which is the minimum between that required to eject the rest of the gas from the halo potential and quench star formation ($f_\star^\mathrm{ej}$) and an upper threshold ($f_\star \sim 1-3\%$) such that $\feff = \min[ f_\star, f_\star^\mathrm{ej}]$. This star formation is assumed to be uniformly distributed over the entire redshift step.\\
 
\item{\it Supernova feedback}: we assume each SNII to produce an energy equal to $E_{51}=10^{51}$ ergs of which a fraction ($f_w$) couples to gas. This work implements a \quotes{delayed SN feedback} scheme that accounts for the mass-dependent lifetimes of stars before they explode as SNII. The snapshots of the N-body simulation scale as the log of the scale-factor, resulting in increasingly longer time-steps with decreasing $z$. So, while at $z\gsim 9$ the delayed SN feedback scheme differs significantly from the instantaneous one, these schemes become increasingly similar with decreasing $z$ until there is effectively no difference at $z\simeq 6$.\\

\item{\it Ionizing photon production}: the intrinsic spectrum of a stellar population sensitively depends on its age ($t$) and metallicity ($Z$). In the interest of simplicity, in this paper, we assume all stellar populations to have a stellar metallicity of $Z= 0.05 ~ \Zsun$. This assumption is justified by studies of high-$z$ Damped Lyman Alpha (DLA) systems which indicate values of a few percent of $\Zsun$ at $z\sim 5$ \citep[e.g.][]{rafelski2012}. We consider two stellar population synthesis models: {\sc Starburst99} \citep{leitherer1999} and {\sc BPASS} \citep{eldridge2017}. The intrinsic UV luminosity $L_{\nu}$ $[\mathrm{erg~s^{-1}~Hz^{-1}~\Msun^{-1}}]$, is quite similar in both models and evolves with time as \citep[see][]{astraeus1}:
\begin{eqnarray}
L_{\nu}(t) &=& \nonumber
 \begin{cases}
  8.24\times10^{20}&\mathrm{for}\ t<4\mathrm{Myr} \\
  2.07\times10^{21}\ \left( \frac{t}{2\mathrm{Myr}} \right)^{-1.33} &\mathrm{for}\ t\geq4\mathrm{Myr}
 \end{cases}
 \label{eq_LUV}
\end{eqnarray}

In this work we use the \textlcsc{Starburst99} model. The \textlcsc{BPASS} model yields the same reionization histories and topologies once the escape fraction of ionizing photons are tuned to yield the observed optical depth \citep[see discussion in Sec. 6 of][]{astraeus1}.\\

\item{\it Patchy UVB and reionization feedback}: in order to simulate (patchy) reionization, we use the \textlcsc{Cifog} code \citep{hutter2018}. This is a MPI-parallelised, semi-numerical reionization code that uses the ionizing emissivities and positions of the underlying galaxy population and the local gas density (on a $512^3$ grid) to yield the photoionization rate, residual \HI fraction and recombination rate of each cell. For galaxies lying in reionized regions, we calculate the fraction of gas mass they can retain after radiative feedback; galaxies in neutral regions are naturally unaffected by reionization feedback. In this work we consider 3 of the cases studied in \citep{astraeus1}: \textit{Early heating}, \textit{Photoionization} and \textit{Jeans mass}. Each UV feedback model has an associated \quotes{characteristic mass} ($M_c$) at which halos can retain half of their gas mass. $M_c$ is minimum for the \textit{Photoionization} model and maximum for the \textit{Jeans mass} model (see Table \ref{table:models}). This is because the \textit{Photoionization} model accounts for gas reacting to an increase in the IGM temperature (to $T_{IGM}=10^4$ K) on a dynamical timescale. However, for the \textit{Jeans mass} model, the IGM is assumed to be heated to $4\times10^4$~K via photoheating upon ionization and the rise in temperature is assumed to translate immediately into a higher Jeans mass. As shown in Table \ref{table:models}, both these models use a constant value of $f_\mathrm{esc}$. Here, we also study an intermediate scenario where $f_\mathrm{esc}$ scales with the gas fraction ejected from galaxies which results in an increase with decreasing halo mass (the \textit{Early heating} model), although we allow gas a dynamical timescale to react to the increase in the IGM temperature to $4\times10^4$~K.
\end{itemize}

\begin{table*}
	\caption{For different redshifts $z$ (column 1), is reported the redshift range ($z_{\rm min}$ and $z_{\rm max}$ in columns 2 and 3, respectively) and the comoving distance $R$ (in units of cMpc) between $z_{min}$ and $z_{max}$ (column 4). The corresponding volumes $V$ (in units of $10^6~{\rm cMpc^3}$) probed by the three surveys (JADES-deep, JADES-medium and WFIRST1) are reported in columns 5-7. Columns 8-10 report the number of sub-volumes $n$ obtained from our simulation box for each survey and redshift, used to quantify the cosmic variance.}
	\centering
	\vspace{2mm}
	\begin{tabular}{c c c c c c c c c c}
		\hline\hline
		$z$ & $z_{\rm min}$ & $z_{\rm max}$ & $R$ & $V$ & $V$ & $V$ & $n$ & $n$ & $n$\\
		& & & & JADES-deep & JADES-medium & WFIRST1 & JADES-deep & JADES-medium & WFIRST1 \\
		\hline
		12.0 & 11.5 & 12.5 & 170.38 & 0.0670 & 0.2767 & 5.2437 & 121 & 25 & 1\\
		10.0 & 9.5  & 10.5 & 218.90 & 0.0796 & 0.3287 & 6.2287 & 144 & 36 & 1\\
		9.0  & 8.5  & 9.5  & 252.54 & 0.0874 & 0.3610 & 6.8397 & 144 & 36 & 1\\
		8.0  & 7.5  & 8.5  & 295.76 & 0.0965 & 0.3986 & 7.5515 & 169 & 36 & 1\\
		7.0  & 6.5  & 7.5  & 352.85 & 0.1071 & 0.4424 & 8.3827 & 169 & 36 & 1\\
		6.0  & 5.5  & 6.5  & 430.95 & 0.1195 & 0.4935 & 9.3502 & 196 & 36 & 1\\
		\hline
		\hline
	\end{tabular}
	\label{table:volumes}
\end{table*}

\begin{figure}
	\centering
	\includegraphics[width=0.95\linewidth]{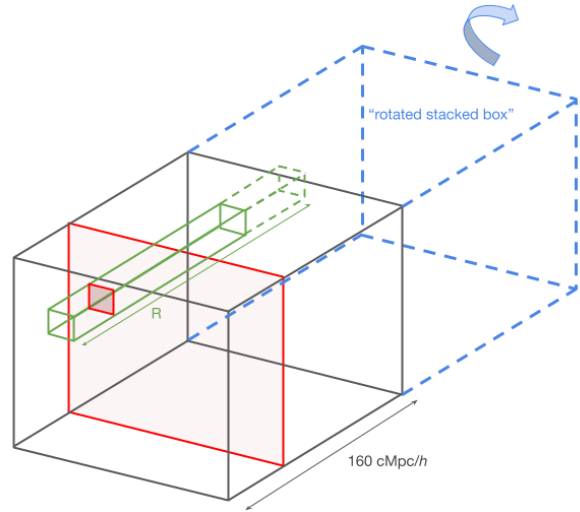}
	\caption{The setup used in this work for computing the sub-volumes used to quantify the cosmic variance. In red, we show the slice of our simulation box reported in Fig. \ref{fig:spatial}, in green is shown the sub-volume having length $R$, i.e., the comoving length corresponding to the survey depth (see Table \ref{table:volumes}). If $R$ is longer than the simulation box (like the example reported in figure), we stack the same box rotated by 180 degrees, i.e. \quotes{rotated stacked box} to the original simulation box (see text for details): in this case, the sub-volume partially overlaps with the stacked box (denoted by the dashed green line).}
		\label{fig:scheme}
\end{figure}

\section{Quantifying cosmic variance}
We now discuss how we calculate cosmic variance in our models and its implications for forthcoming surveys. As shown in Appendix B of \citet{astraeus1}, we find that our model converges for halos with a DM mass of $M_h \geq 10^{8.6} \msun$ (corresponding to halos with at least 50 particles). In this work, we limit all analyses to such resolved halos.

\subsection{Methodology and surveys considered}
\label{sec_surveys}
In this work, we focus on the following three forthcoming surveys, planned for the JWST and NGRST:\\
({\it i}) JWST JADES-deep survey\footnote{\scriptsize https://issues.cosmos.esa.int/jwst-nirspecwiki/display/PUBLIC/Overview}: area of 46 arcmin$^2$;\\
({\it ii}) JWST JADES-medium survey: area of 190 arcmin$^2$;\\
({\it iii}) WFIRST1 survey\footnote{\scriptsize https://roman.gsfc.nasa.gov/science/WFIRSTScienceSheetFINAL.pdf}: area of 1 deg$^2$.

For each survey, at each redshift $z$, we calculate the comoving volume $V$ corresponding to the survey area enclosed between the minimum and maximum redshift ($z_{\rm min}$ and $z_{\rm max}$) values reported in Table \ref{table:volumes}. Further, $R$ is the comoving distance between $z_{\rm min}$ and $z_{\rm max}$ as shown in Fig. \ref{fig:scheme}. If $R$ is longer than the simulation box (i.e. for $z < 9$), we stack the same box rotated by 180 degrees around the $R$ axis (\quotes{rotated stacked box}) to the original simulation box as shown in the same figure. The number of sub-volumes ($n$) obtained from our simulation\footnote{The transverse length for each of our sub-volumes is $\sqrt{V / R}$. This yields $n = [160 h^{-1} \rm cMpc \cdot \sqrt{R/V}]^2$, where the square brackets denote the integer part, and $R$ and $V$ are in units of cMpc.}, from $z \sim 6-12$, for each survey are reported in Table \ref{table:volumes}.

We calculate the cosmic variance in the evolving HMF, UV LF and SMF as \citep{Driver2010}:

\begin{equation}
\zeta (\%) = 100 \cdot \sigma_{var},
\label{eq:sigma}
\end{equation}
where $\sigma_{var}$ quantifies the uncertainties in excess to Poisson shot noise and is defined as \citep[e.g.,][]{Trenti2008}:

\begin{equation}
\sigma_{var} = \sqrt{\frac{\left<N^2\right> - \left<N\right>^2}{\left<N\right>^2} - \frac{1}{\left<N\right>}}.
\label{eq:variance}
\end{equation}
Here, $\left<N\right>$ is the mean galaxy number across all the sub-volumes and $N$ the number of galaxies in each sub-volume. The analysis performed in this work is done considering the mass and luminosity functions measured at single snapshots of our simulation (namely the mean of the observed redshift ranges reported in the first column of Table \ref{table:volumes}) instead of constructing observational light cones. We have checked that constructing a light cone does not substantially change our results. To this end we used the few stored outputs in the corresponding redshift intervals and stacked parts of the outputs together to mimic roughly a light cone.

To allow interested readers to compute cosmic variance for different redshifts, redshift intervals and survey areas, we also provide a public \textlcsc{Python} user-friendly tool\footnote{https://github.com/grazianoucci/cosmic\_variance}.

\begin{figure*}
	\centering
	\includegraphics[width=0.98\linewidth]{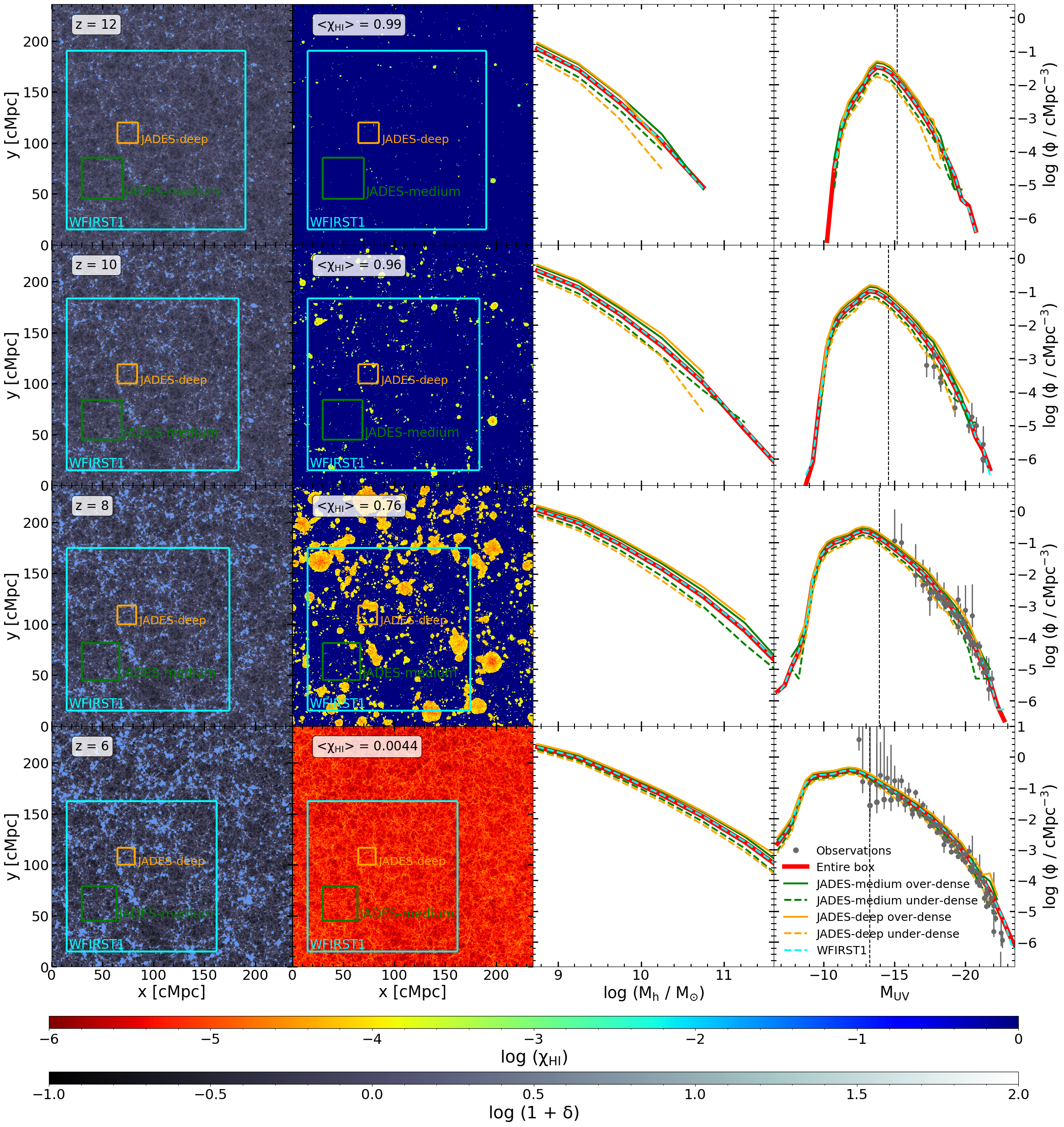}
	\caption{A slice through our simulation box for the \textit{Photoionization} model with columns (from \textit{left} to \textit{right}) showing : the density fields ($1 + \delta$), the neutral hydrogen fraction ($\chi_{\rm HI}$), the evolving halo mass function (HMF) and the UV luminosity function (UV LF) at redshifts 12 to 6 as marked in the left-most panels; the values of the density and ionization fraction can be read using the color-bars at the bottom of the figure. The second column reports also the average neutral hydrogen fraction in the simulation ($<\chi_{\rm HI}>$). The squares in the left-two columns denote the size corresponding to the survey areas considered in this work as marked (JADES-deep, JADES-medium, WFIRST1). Blue dots in the left-most panel represent galaxies with $\muv < -15$ with their size proportional to their UV luminosity. In the right-two columns, continuous and dashed lines show the UV LF and HMF for the most over-dense ($\left<1 + \delta\right> \sim 1.1$ as reference, for the JADES areas) and under-dense sub-volumes ($\left<1 + \delta\right> \sim 0.9$ as reference, for the JADES areas), respectively. Red continuous lines show the UV LF and HMF computed over the entire box. The point with error bars represent the observational UV LF data at: $z = 10$ \citep{oesch2013,bouwens2015b,bouwens2016b,oesch2018}, $z = 8$ \citep{bradley2012,mclure2013,schenker2013,atek2014,bouwens2015b,finkelstein2015,livermore2017} and $z = 6$ \citep{willott2013,bouwens2015b,bowler2015,finkelstein2015,bouwens2017,livermore2017,atek2018}. The grey dashed lines in the last column mark the $\muv$ limit that might be affected by the resolution limit of the underlying N-body simulation \citep{astraeus1}.}
	\label{fig:spatial}
\end{figure*}

We compare the above-mentioned surveys with our simulation snapshots at $z \sim 6-12$ (with $\Delta z=2$) in Fig. \ref{fig:spatial}. At each redshift, we show the LBGs (with $\muv <-15$) distribution embedded in the large-scale density and the ionization fields (calculated on a $512^3$ grid) along with the resulting HMF and UV LF for each survey. The density of each cell is calculated as $1 + \delta = \rho/\rho_{\rm crit}$, where $\rho$ and $\rho_{\rm crit}$ are the dark matter density in the grid cell and the critical density at that $z$, respectively. The volume-weighted neutral hydrogen fraction is directly obtained from the results of \textlcsc{CIFOG}. We briefly discuss where LBGs (with $\muv <-15$) lie in terms of the density and neutral fractions; this is explored in more detail in Sec. \ref{sec:environment}. Firstly, with $M_h \gsim 10^{9.5} \msun$ and $10^{8.8} \msun$ at $z \sim 6$ and 12, respectively, these LBGs typically lie in over-dense regions (with $1+\delta \sim 1-100$) as shown in the first column of Fig. \ref{fig:spatial}. Further, in the \textit{Photoionization} model, all galaxies have a constant value of $f_\mathrm{esc}=0.215$. This naturally results in the most massive/luminous star forming galaxies being able to create the largest ionized regions around themselves. As shown in column 2 of the same figure, this means that the densest regions are reionized first, as expected for the inside-out reionization scenario. We also see the evolution of the volume filling fraction of ionized hydrogen ($Q_{II}$) in column 2. As shown, the IGM is effectively neutral at $z\sim 12$ where $Q_{II}\sim 0$, reaches a mid-point at $z \sim 7.2$ and reionization is over ($Q_{II}\sim 1$) by $z\sim 6$.

As shown in column 3 of Fig. \ref{fig:spatial}, the HMF evolves positively both in terms of normalisation and the halo mass extent with decreasing redshift. Further, the amplitude of the HMF correlates with the underlying density-field: the most under-dense ($\left<1 + \delta\right> \sim 0.9$ as reference, for the JADES areas) and over-dense ($\left<1 + \delta\right> \sim 1.1$ as reference, for the JADES areas) sub-volumes show the lowest and highest amplitude of the HMF, although its shape remains unchanged. The number of sub-volumes involved in the computation of the continuous and dashed lines are reported as $n$ in Table \ref{table:volumes}. Given its smallest area, the JADES-deep survey shows the largest variation in the amplitude of the HMF in over/under-dense regions with the variance decreasing with an increase in the survey area (that averages over a larger range in halo bias). Having only 1 sub-volume corresponding to the survey area for WFIRST1, we can not calculate the variance. The HMF naturally shows an increasing variance with halo mass given their decreasing number densities. Finally, since halos of a given mass become increasingly less biased with decreasing redshift, their variance decreases with decreasing redshift too.

The same trends are seen for the UV LF (column 4 in Fig. \ref{fig:spatial}) where the most under/over-dense regions yield the minima and maxima. Again, the variance increases with decreasing magnitude given the rarity of these sources. Finally, at a given UV magnitude, the variance decreases with decreasing redshift.

Interestingly, although the model is tuned to match the UV LF over the entire box, given the error bars on observational results, as of now, all of our sub-volumes yield UV LFs in agreement with the available observations at $z =6-10$.

\begin{figure}
	\centering
	\includegraphics[width=1.0\linewidth]{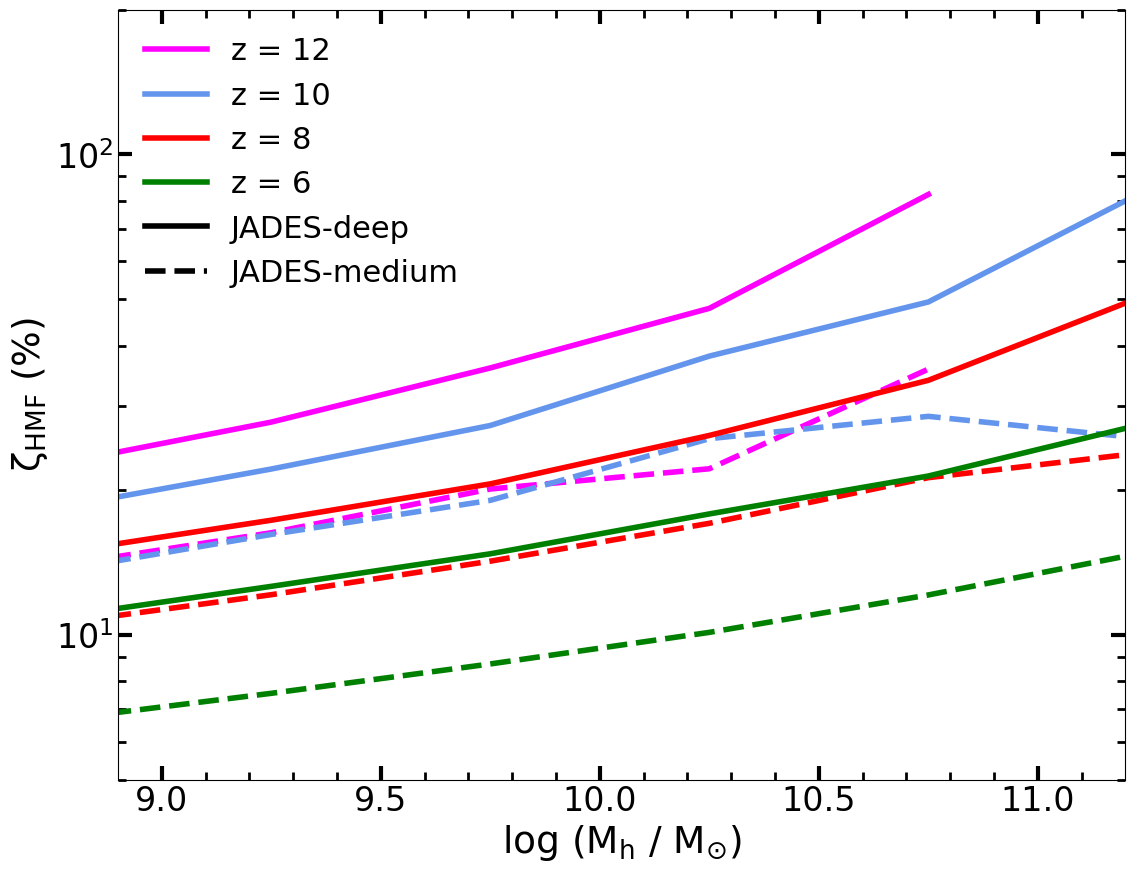}
	\caption{The cosmic variance as a function of halo mass at different redshifts for the JADES-deep and JADES-medium survey volumes, as marked.}
	\label{fig:cosmic_variance_hmf}
\end{figure}

\subsection{The origin and quantification of cosmic variance}
\label{origin}
We start by quantifying the cosmic variance in the HMF for $z \sim 6-12$ as shown in Fig. \ref{fig:cosmic_variance_hmf}. The cosmic variance increases with halo mass for both the JWST surveys considered here. At a given $M_h$ the variance increases with increasing $z$ given the decrease in the comoving areas probed. As noted, our simulation area is of the order of the WFIRST1 survey and so we can not reliably calculate the variance for it with the formalism described in eqs. (\ref{eq:sigma}) and (\ref{eq:variance}). Starting with the JADES-deep survey, the variance increases from about 20\% to $\sim$ 80\% as $M_h$ increases from $10^9$ $\Msun$ to $10^{11}$ $\Msun$ at $z = 10$. Further, while at the lowest halo masses ($M_h \sim 10^9 \ \Msun$) the variance only changes by a factor of $\sim$ 2 between $z = 6-12$, it increases by $\sim$ 4 for the most massive halos ($M_h \sim 10^{10.5}\Msun$) between the same redshift range given the rarer/biased fluctuations required to form the latter halos. As expected, the cosmic variance naturally decreases with an increasing survey area: as the survey area increases by a factor 4 from JADES-deep to JADES-medium, the cosmic variance decreases by a factor $\sim$ 2-4 for the mass range considered.

\begin{figure*}
	\centering
	\includegraphics[width=0.49\textwidth]{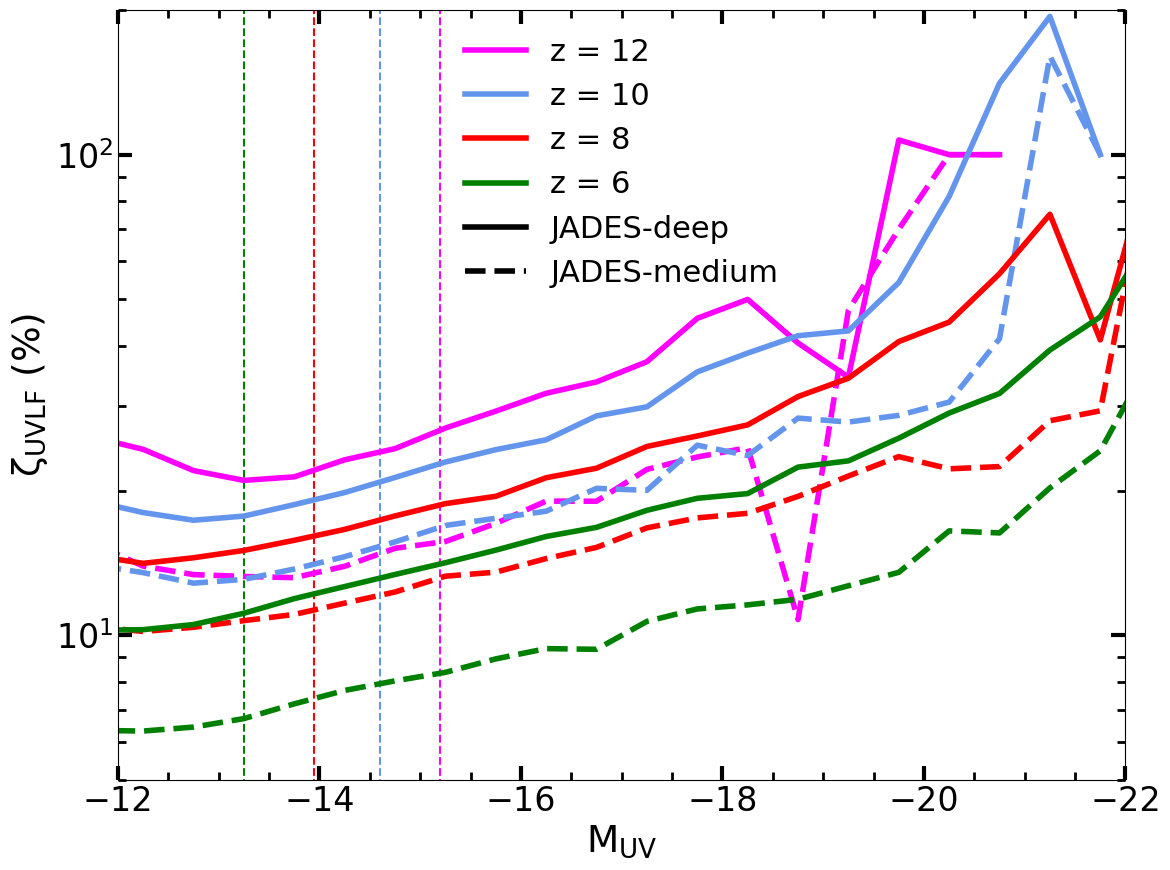}
	\includegraphics[width=0.49\textwidth]{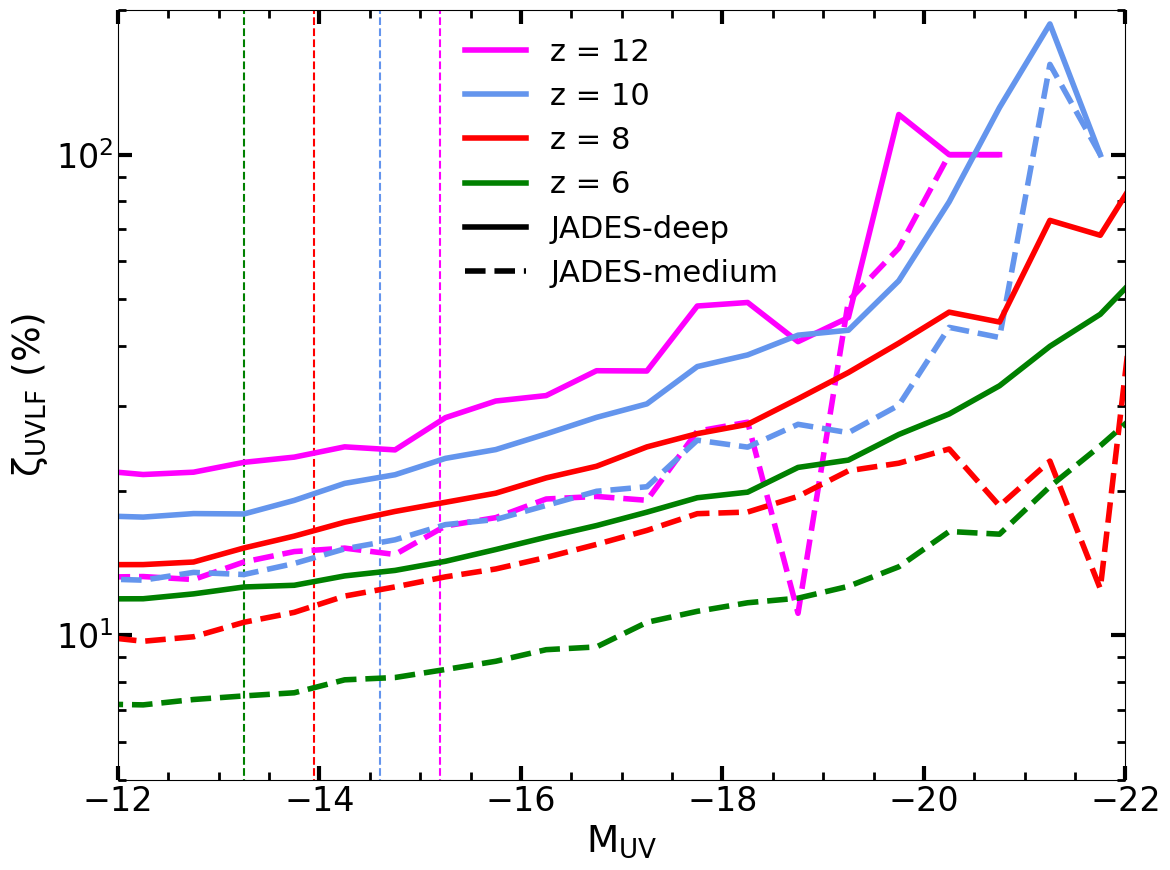}
	\caption{The cosmic variance as a function of UV magnitude at different redshifts for the JADES-deep and JADES-medium survey volumes, as marked. The \textit{left} and \textit{right} panels show results for the \textit{Photoionization} and \textit{Jeans mass} models, respectively. Vertical dashed lines indicate the $\muv$ limit that might be affected by the resolution limit of the underlying N-body simulation \citep{astraeus1} color-coded by redshift.}
	\label{fig:cosmic_variance_uvlf}
\end{figure*}

Furthermore, we study the cosmic variance in the UV LF for \textit{Photoionization} and \textit{Jeans mass} models, the results of which are shown in Fig. \ref{fig:cosmic_variance_uvlf} for the JADES-deep and JADES-medium surveys, with vertical dashed lines indicating the limit below which results are subject to the non-convergence of galactic properties. Starting with the \textit{Photoionization} model, we find the variance to be minimum (i.e., $\zeta \lsim 30\%$) between $\muv \sim -12$ and $\muv \sim -15$ at $z \sim 6-12$. The variance then increases in both directions around this range: for the JADES-deep survey, $\zeta > 30\%$ for $\muv \sim -17.5$ at $z \sim 12$ and for a lower value of $\muv \sim -20$ by $z=6$. As expected, the JADES-medium survey reaches the same variance at lower magnitudes of $\muv \sim -19 ~(-22)$ at $z =12~ (6)$. On the other hand, while $\zeta < 10\%$ for $\muv \sim -12$ at $z = 6$ for both JADES surveys, it increases at earlier times (up to 25\% at $z=12$). This is the result of the stochastic star formation and the quick decay of UV luminosity with time in low-mass halos that results in low-mass galaxies of a given mass being distributed over a wide $\muv$ range. Reionization effects only kick in at $\muv\gsim -9$ for this model and are therefore not the key driver of the variance seen at the low-luminosity end. Finally, for a given $\muv$ the variance decreases by a factor of about 2 between the JADES-deep and JADES-medium survey, almost independent of $\muv$, given the larger area probed by the latter. We then explore the impact of our strongest reionization model (\textit{Jeans mass}), where the gas in ionized regions is instantaneously heated to $4\times 10^4$ K and the gas density instantaneously reduced, leading to an immediate increase of the Jeans mass (and hence characteristic mass), in the right panel of Fig. \ref{fig:cosmic_variance_uvlf}. As expected, also this reionization model shows the same variance for bright ($\muv\lsim -14$) galaxies for both JWST JADES surveys. The impact of reionization feedback on their low-mass progenitors has essentially no effect on the gas content and assembly of these high-mass systems. This model shows approximately the same variance as the \textit{Photoionization} model.

\begin{figure*}
	\centering
	\includegraphics[width=1.05\linewidth]{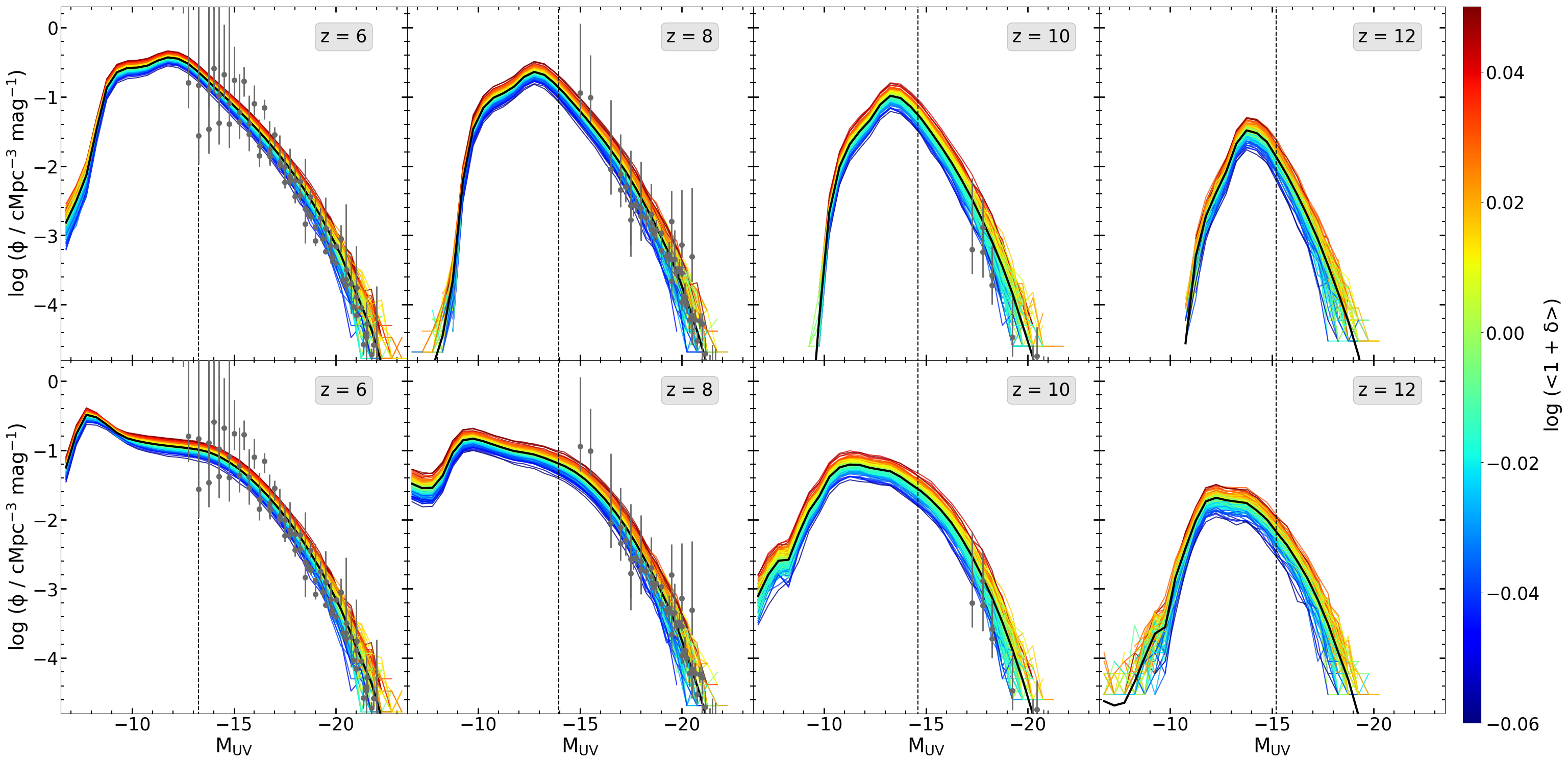}
	\caption{UV LFs for the different sub-volumes in our simulation box corresponding to the JADES-deep survey volume at different redshifts color-coded by the corresponding average overdensity in each sub-volume. The upper and lower panels show results for reionization feedback in the \textit{Photoionization} and \textit{Jeans mass} models, respectively (see Table \ref{table:models} for details). Vertical dashed grey lines indicate the $\muv$ limit that might be affected by the resolution limit of the underlying N-body simulation \citep{astraeus1}.}
	\label{fig:colorcoded_density_uvlf}
\end{figure*}

In Fig. \ref{fig:colorcoded_density_uvlf} we study the UV LFs in these two different reionization models as a function of over-density ($1 + \delta$) to understand the reason for this variance. We find that most of the cosmic variance is completely driven by the underlying density field. Fig. \ref{fig:colorcoded_density_uvlf} clearly shows that although the number density of galaxies scales with the density of their environments, the shape of the UV LF is quite independent of the environment. This essentially shows galaxy assembly to be mostly driven by \quotes{local} processes with almost no dependence on the environment. Quantitatively, a factor of 1.3 change in the over-density causes a variance less than 25\% at $\muv \sim -14$ for all $z$ and UV feedback models. $\zeta$ increases to 30\% (100\%) at $\muv \sim -20$ at $z = 6$ ($z = 12$) for the JADES-deep survey with the variance being a factor $\sim 1.6$ lower for the JADES-medium survey. However, as noted above, as of now, all of the over-densities probed are within current observational uncertainties.

\begin{figure}
	\centering
	\includegraphics[width=1.0\linewidth]{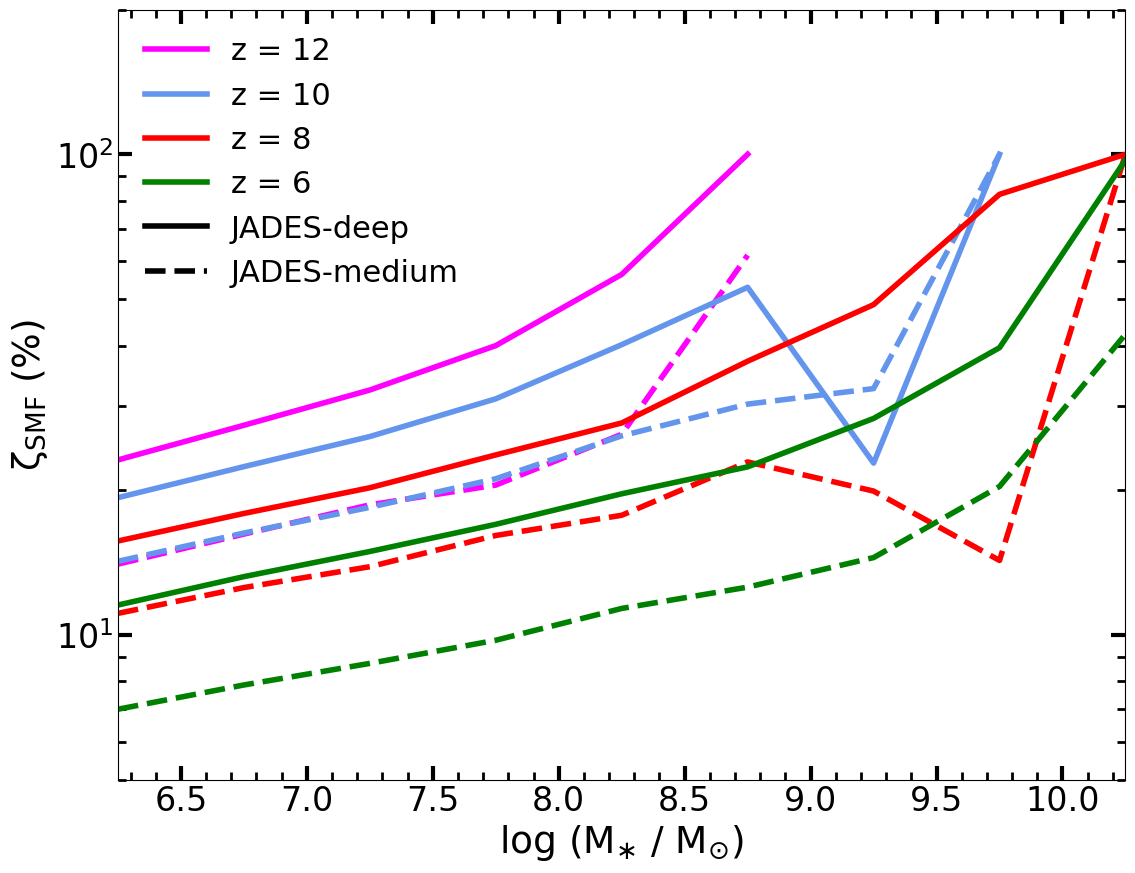}
	\caption{The cosmic variance as a function of stellar mass for the \textit{Photoionization} model as a function of stellar mass at different redshifts for the JADES-deep and JADES-medium survey volumes, as marked.}
	\label{fig:cosmic_variance_smf}
\end{figure}

\begin{figure*}
	\centering
	\includegraphics[width=1.0\linewidth]{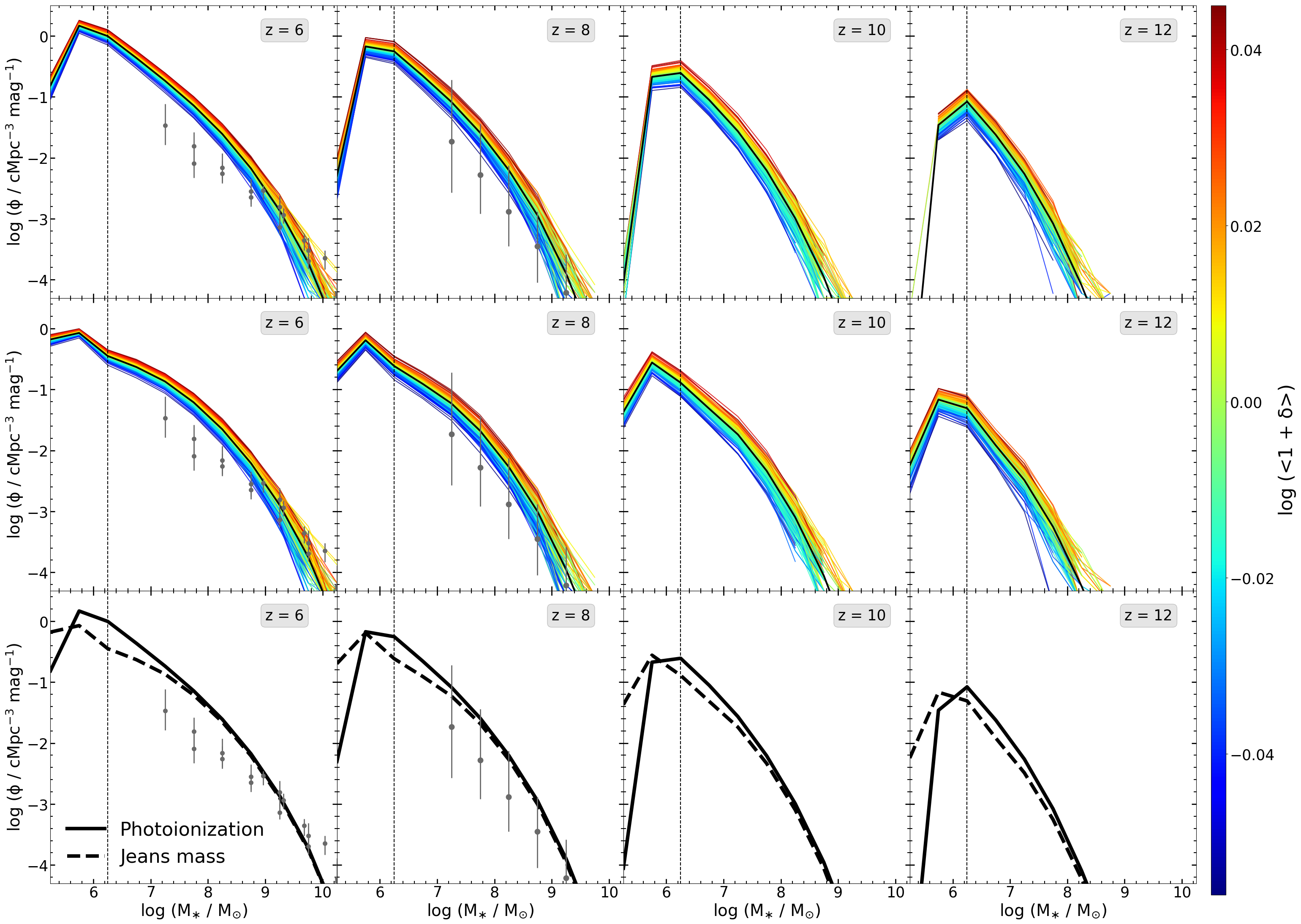}
	\caption{SMFs for the different sub-volumes in our simulation box corresponding to the JADES-deep survey volume at different redshifts color-coded by the corresponding average overdensity in each sub-volume. The \textit{upper} and \textit{middle} panels show results for reionization feedback in the \textit{Photoionization} and \textit{Jeans mass} models, respectively (see Table \ref{table:models} for details). For a comparison, the \textit{lower} panels show the SMF averaged over the entire simulation box for the two models. The points with error bars represent the observational SMF data at $z = 6$ \citep{gonzalez2011,duncan2014,song2016}, and $z = 8$ \citep{song2016}. For a comparison between the models and observations see Sec. 3.2 in \citet{astraeus1}. Vertical dashed grey lines indicate the stellar mass limit that might be affected by the resolution limit of the underlying N-body simulation \citep{astraeus1}.}
	\label{fig:colorcoded_density_smf}
\end{figure*}

We show the cosmic variance as a function of stellar mass in Fig. \ref{fig:cosmic_variance_smf} for the \textit{Photoionization} model; we note that the results here only differ from the \textit{Jeans mass} model at the percent level. As seen from this figure, the variance tends to increase with stellar mass (given their increasing rarity/bias) for all redshifts for both the JWST survey considered. Further, since a given stellar mass samples an increasing bias with increasing redshift, it naturally leads to an increase in the cosmic variance. Starting with the JADES-deep survey, we find $\zeta \lsim 15\%$ for $M_* <10^8 \msun$ at $z=6$. Galaxies of a much lower stellar mass $M_* \sim 10^{6.3}\msun$ show the same variance by $z = 8-10$. For a redshift as high as $z=12$, $\zeta > 20\%$ even for galaxies as low-mass as $M_* \sim 10^{6.3}\msun$. The variance in the stellar mass reaches a value of 100\% for $M_* \sim 10^{10.2}~ (10^{8.5})\msun$ at $z= 6 ~ (12)$. We note that the lowest stellar masses probed as of now correspond to $M_* \sim 10^{7.2}\msun$ at $z=6-8$ where $\zeta<20\%$, even with a survey as focused as the JADES-deep.

The stellar mass function for the JADES-deep survey, color-coded by the over-density in each field, is shown in Fig. \ref{fig:colorcoded_density_smf}. As expected from the hierarchical structure formation model, the amplitude of the SMF decreases with increasing stellar mass and increasing redshift; additionally, the mass range of the SMF decreases with increasing redshift. Analogous to the UVLF, for both the \textit{Photoionization} and \textit{Jeans mass} models, the amplitude of the SMF also scales with the over-density of the region sampled, showing that most of the cosmic variance is driven by the underlying density field. Moreover, at the most massive end, a decrease in the over-density by a factor of about 1.3 leads to a maximum halo mass that is lower by a factor of about 5.

We find that the SMFs from the \textit{Photoionization} and \textit{Jeans mass} models are indistinguishable for $M_* \gsim 10^{7.5}\msun$, corresponding to $M_h \gsim 10^{10}\msun$, where the impact of UV feedback is mostly irrelevant. However, the shape of the SMFs is clearly different between these two models for lower stellar masses. For $M_* \sim 10^{6.4}\msun$, the amplitude of the SMF is a factor 3 lower in the Jeans mass model (compared to the \textit{Photoionization} model) given its much larger suppression of gas mass in low-mass halos.

Finally, we have computed that, in order to have the HMFs, UV LFs and SMFs converging towards those computed over the entire simulation box (i.e., the difference between the density functions in every sub-volume is $<$ 0.05 dex from the ones in the entire simulation box), an area of $\sim$ 1000 arcmin$^2$ is needed. Therefore this could represent a survey area to minimize cosmic variance, specially at the low-mass end. This area can also be built combining several pointings. Although each single pointing could be affected by a relatively large cosmic variance, the average UV LF or SMF resulting from their combination will be substantially independent of the underlying density field.

\citet{Bhowmick2019} also found that a survey area of 10 deg$^2$ (i.e., WFIRST10) will have a cosmic variance ranging of about 6-10\%, while upcoming JWST surveys (with areas up to $\sim$ 100 arcmin$^2$) will have cosmic variance ranging from $\sim 20-40\%$ for fainter galaxies ($H < 29$ mag, i.e., $\muv \sim -18$ at $z \sim 8$) to $\sim 50-100\%$ for brighter objects ($H < 25$ mag, i.e., $\muv \sim -22$ at $z \sim 8$) at $z \sim 7.5-8$. This is perfectly in agreement with our estimates, and the scatter in their apparent magnitude function (i.e., the number density of galaxies having a certain $H$ mag) shows the same trend we obtained (i.e., $\sim$ 0.5 dex for 10 arcmin$^2$ at $H \sim 27$ ($\muv \sim -20$ at $z \sim 8$) \footnote{Given the wide range of mass-to-UV light ratios, the conversions between the H band magnitude and $\muv$ are meant to be rough estimates.}

\subsection{Quantifying the UV LF faint end slope - constraints on reionization}
The faint end slope of the UV LF ($\alpha$) is shaped by two types of feedbacks: \quotes{internal} feedback from SNII that can heat or blow-out a significant fraction of gas (or even all of it) from the small potentials of low-mass halos \citep[e.g.,][]{maclow-ferrara1999} and \quotes{external} feedback from a rising UVB that can photo-evaporate gas or prevent gas accretion onto low-mass halos, thereby suppressing further star formation \citep[e.g.,][]{barkana1999,shapiro2004,petkova2011, finlator2011, hasegawa2013}. SNII feedback effectively depends on the ratio of the SNII energy and the binding energy. However, due to patchy reionization, the strength of the UVB and hence UV feedback, vary as a function of spatial position and time. In ionized regions, the faint-end slope of the UV LF becomes shallower due to the decreasing star formation efficiencies of low-mass halos. Indeed, as pointed out in \citet{choudhury2019}, the faint-end slope in different JWST fields could be an extremely powerful tool to study the differential feedback impact of patchy reionization on galaxy formation.

\begin{figure*}
	\centering
	\includegraphics[width=1.0\linewidth]{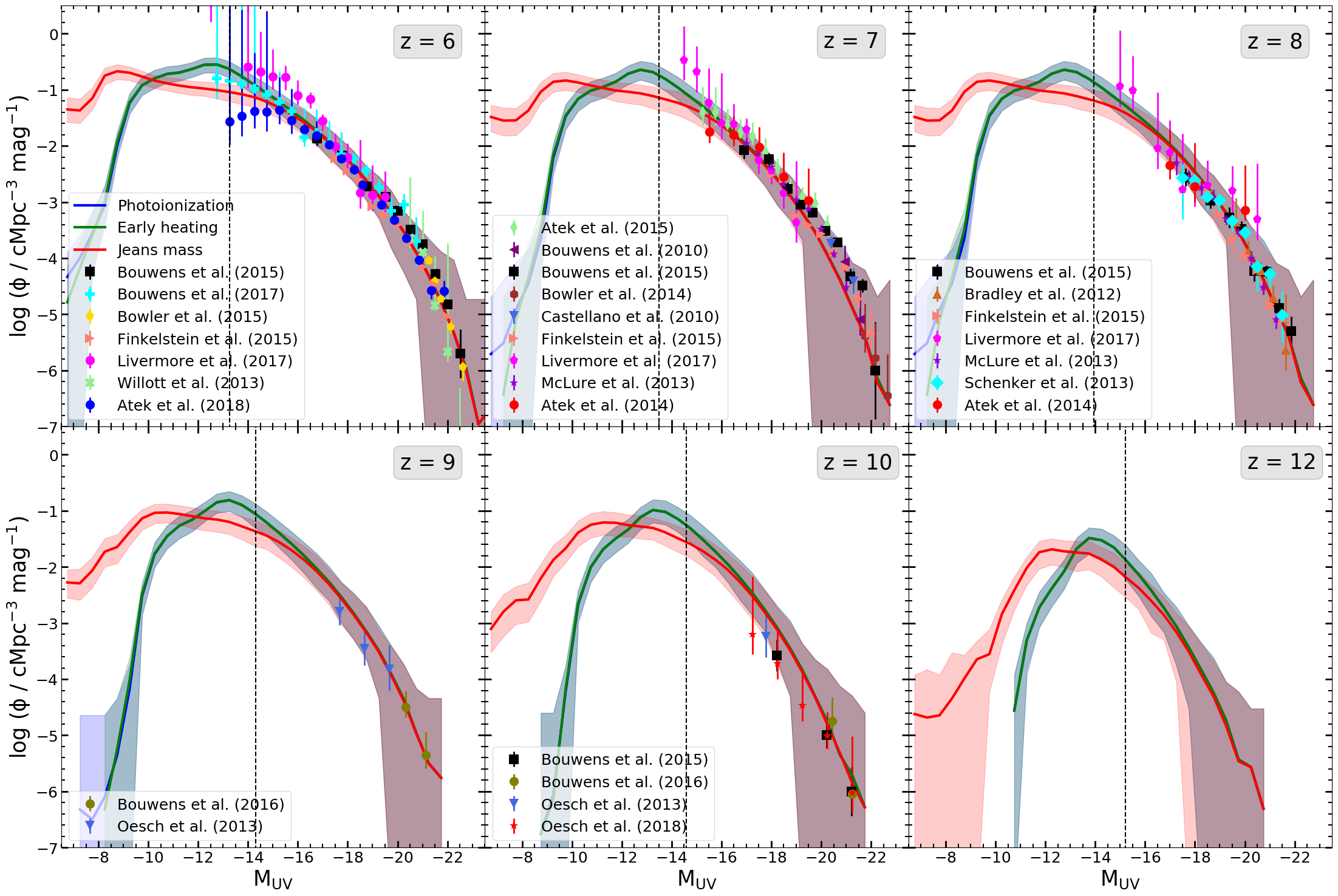}
	\caption{UV LFs at various redshifts for the different feedback models considered in this work (see Table \ref{table:models} for details) for the JADES-deep survey. Continuous line correspond to the average UV LF in our simulation box, while the shaded area delimits the lower and upper UV LF obtained in the sub-volumes. Points show observed UV LFs, with the reference marked in each panel. Vertical dashed grey lines indicate the $\muv$ limit that might be affected by the resolution limit of the underlying N-body simulation \citep{astraeus1}.}
	\label{fig:uvlf}
\end{figure*}

We start by studying the UV LFs in the JADES-deep field survey given its ability to probe the UV LF faint end\footnote{JADES-deep has NIRCam limits at mag $\sim$ 29.8 corresponding to $\muv \sim -16.9$ at $z \sim 6$, $\muv \sim -17.3$ at $z \sim 8$, and $\muv \sim -18$ at $z \sim 12$. The JADES-medium survey is 1 magnitude shallower.} for the impact of reionization feedback as shown in Fig. \ref{fig:uvlf}. We chose JADES-deep in order to explore a case with a higher cosmic variance so as to be able to study the maximum uncertainties in the faint end slope. We study three reionization models: \textit{Photoionization}, \textit{Jeans mass} and \textit{Early heating}. We remind the reader that while the first two models use a constant value for $f_{esc}$, the last is the only one where $f_{esc}$ increases with decreasing halo mass \citep{astraeus1}. We start by showing the UV LFs expected for these three UV feedback models, along with the expected cosmic variance, in Fig. \ref{fig:uvlf}. We note that, as of now, all three models are in agreement with observational data at $z=6-10$ within error bars. Further, cosmic variance becomes of the order of 100\% at $\muv \sim -19$ given the small volumes probed by the JADES-deep fields. At the faint-end, UV feedback leads to a flattening in the UV LF at $\muv \sim -14.5$ at $z \sim 6-8$ for the \textit{Jeans mass} model where the gas mass of low-mass halos is instantaneously suppressed by reionization feedback. The \textit{Photoionization} and \textit{Early heating} models are extremely similar for what concerns the UV LF and show a decrease at lower magnitudes of $\muv \sim -13$ at all redshifts \citep[for a complete discussion see][]{astraeus1}.

\begin{figure}
	\centering
	\includegraphics[width=0.95\linewidth]{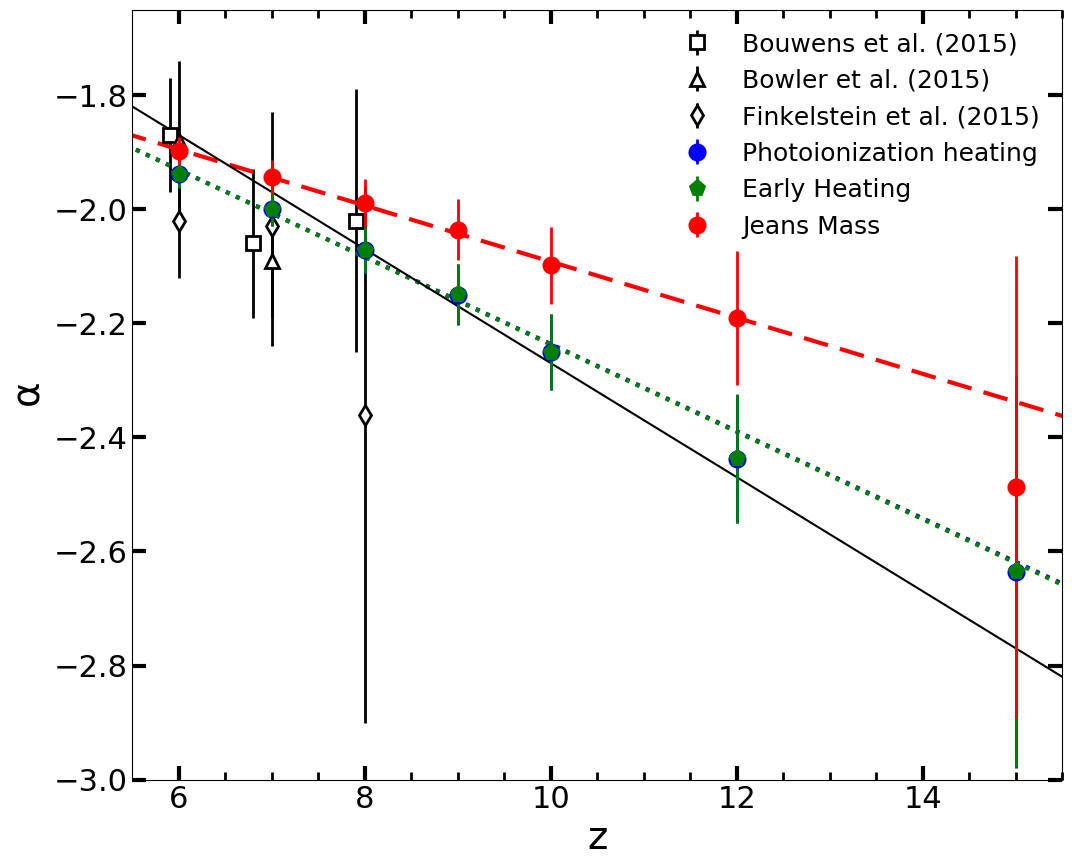}
	\caption{Evolution in faint-end slope of the UV LF ($\alpha$) over the redshift range $6 < z < 12$ (we performed the fit for $\muv < -15$) for the different models considered in this work compared with the results in literature \citep{bouwens2015b,bowler2015,finkelstein2015}. Orange, green and blue dashed lines show the linear fitting functions for the \textit{Photoionization}, \textit{Early heating} and \textit{Jeans mass}, respectively. Black dashed line shows instead the fitting function derived by \citet{bouwens2015b} (see text for details).}
	\label{fig:alpha}
\end{figure}

We now study the redshift evolution of the Schechter function faint-end slope $\alpha$ for $\muv < -15$, along with the cosmic variance, in more detail as shown in Fig. \ref{fig:alpha}. Tracking the steepening of the underlying HMF, the faint-end UV LF slope steepens too with redshift for all models. The value of $\alpha$ is effectively the same (within uncertainties\footnote{$\alpha$ and its uncertainties are derived fitting the UV LF with a full Schechter function for $\muv < -15$, with the errors given by the cosmic variance at each $\muv$.}) in the \textit{Photoionization} and \textit{Early heating} models at all redshifts and evolves as 
\begin{equation}
\alpha(z) = (-1.931 \pm 0.007) + (-0.076 \pm 0.004) (z - 6)
\end{equation}

With its faster suppression of the gas mass, and hence star formation rate and UV luminosity, the \textit{Jeans mass} model shows a shallower redshift evolution of $\alpha$ such that 
\begin{equation}
\alpha(z) = (-1.896 \pm 0.002) + (-0.049 \pm 0.002) (z - 6)
\end{equation}

Within error bars, these slopes are in accord with the observationally inferred evolution of $\alpha$ presented in \citet{bouwens2015b} who assumed the Schechter function parameters to vary linearly with redshift to find:
\begin{equation}
\alpha(z) = (-1.87 \pm 0.05) + (-0.10 \pm 0.03) (z - 6)
\end{equation}

As seen from this plot, even accounting for cosmic variance for $z \geq 9$ (i.e., where the error bar do not overlap), the value of $\alpha(z = 9) = -2.040 \pm 0.005$ in the \textit{Jeans mass} is distinguishable from that in the \textit{Photoionization} and \textit{Early heating} models $\alpha(z = 9) = -2.159 \pm 0.019$. Therefore, integrating down to $\muv = -15$ at $z\sim 9-12$, a reasonable limit for the JWST, would be sufficient to differentiate between \textit{Photoionization} and the \textit{Jeans mass} models.

\section{The environments of LBGs and evolution in EoR}\label{sec:environment}
\begin{figure*}
	\centering
	\includegraphics[width=1.0\linewidth]{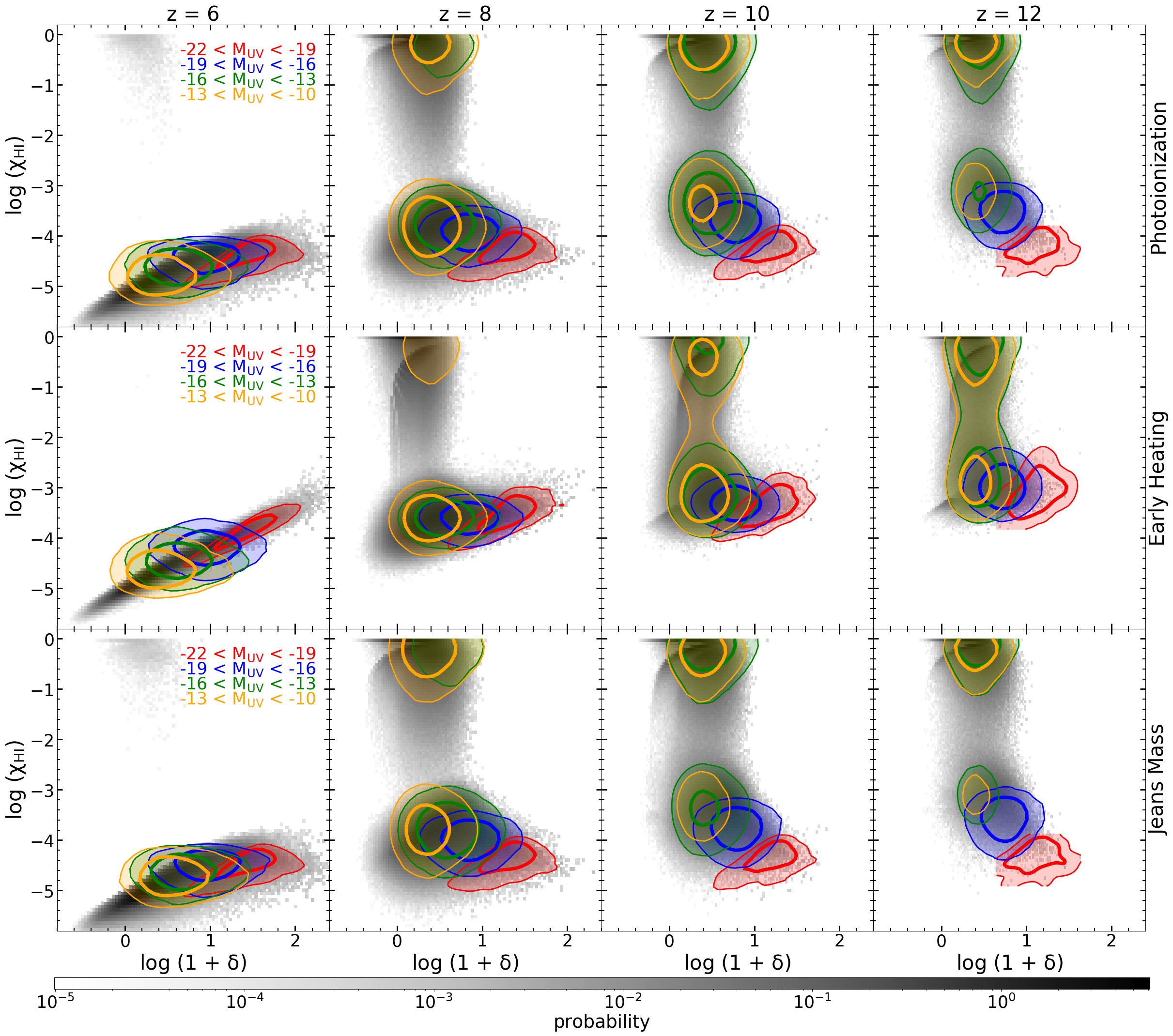}
	\caption{Probability density distribution of the IGM hydrogen neutral fraction $\chi_{HI}$ versus gas over-density (1 + $\delta$) in our simulated box for three different models considered in this work (see Table \ref{table:models}): from top to bottom \textit{Photoionization}, \textit{Early Heating}, \textit{Jeans Mass}. In each panel, the points show the probability distribution function for the cells (with a volume of $[0.3125 h^{-1} {\rm cMpc}]^3$) in the simulation box; the color bar shows the probability distribution. The thick and thin contours show the contours occupied by 68\% and 96\% of galaxies, respectively, in the magnitude bins marked in the top-most left panel.  }
	\label{fig:densities}
\end{figure*}

We now study the environments of LBGs in the EoR, specifically focusing on the over-densities and ionization fractions of the regions they are embedded in as shown in Fig. \ref{fig:densities} \citep[see also][]{hutter2017}. We discuss our results for three UV feedback models: \textit{Photoionization}, {\it Early heating} and {\it Jeans mass}.

Starting with the \textit{Photoionization} model, most of the simulation volume is neutral at $z \sim 10-12$. Here, 68\% of low-luminosity ($\muv \sim -13 - -10$) galaxies with $M_h \sim 10^{8-9}\msun$ occupy slightly over-dense ($1 + \delta \sim 1.5-4$) and neutral regions ($\chi_{HI}>0.4$). In addition to such field galaxies, we also find a fraction of such faint (clustered) galaxies that lie in extremely ionized regions ($\chi_{HI} <10^{-2.6}$). With larger halo masses ($M_h \sim 10^{9.1-10.1}\msun$), $\muv \sim -16$ to $-19$ LBGs occupy more over-dense ($1 + \delta \sim 2.5-25$) and fully ionized ($\chi_{HI}<10^{-3}$) regions. Finally, the most luminous ($\muv \sim -19$ to $-22$) and massive ($M_h \sim 10^{10.1-11}\msun$) LBGs occupy the most over-dense ($1 + \delta \sim 5-50$) and highly ionized ($\chi_{HI}<10^{-4}$) regions, hinting at their longest star formation histories. By $z =8$, a large fraction of the cells in the simulation are ionized. As noted in Sec. \ref{origin}, the faintest galaxies show a negative luminosity evolution as their stellar populations age and very few new stars form due to SNII feedback. Most LBGs now lie in highly ionized regions ($\chi_{HI} <10^{-3}$) with only a small fraction of the faintest isolated LBGs still lying in neutral regions ($\chi_{HI}>0.1$) as a result of their low star formation rates. Finally, reionization finishes by $z=6$ at which point we find a positive correlation between $1+\delta$ and $\chi_{HI}$, except for the most over-dense cells, driven by their high recombination rates \citep[following the relations modelled in \textlcsc{CIFOG}, e.g.,][]{hutter2018}. At this point, all the LBGs studied here lie in completely ionized regions ($\chi_{HI} <10^{-4}$) with the over-density values scaling from $\sim 0.6$ for the faintest galaxies to $\sim$ 100 for the brightest objects.

The ionization fields for the \textit{Early Heating} model differ strongly from the other two models especially in the initial stages of reionization ($z >10$). This is because in this model, the ionizing escape fraction $\fesc$ decreases with increasing halo mass, which results in low-mass galaxies providing a larger fraction of ionizing photons compared to the other two models. In this case, $\muv =-10$ to $-13$ galaxies lie in regions occupying a large range in ionization such that $\chi_{HI} \sim 1-10^{-3.5}$ as early as $z =10-12$. This is because in this case faint galaxies can partially ionize their cells; these cells would be mostly neutral in all the other feedback models considered. Interestingly, as a result of their large output of \HI ionizing photons, a larger bulk of such low-luminosity galaxies lie in ionised regions, compared to the other two models. By $z=8$, LBGs occupy very similar environments, both in terms of density and neutral fraction, as in the other two models. However, we see a mild upturn of the contours at a value of $\chi_{HI} \sim 10^{-3}$ in the simulation cells with $1+\delta \gsim 3$ which is also reflected in the distribution of the most massive galaxies. This upturn in the $1+\delta-\chi_{HI}$ relation (compared to the downturn in the other two models) is driven by the lower escape fraction of \HI ionizing photons from the most massive galaxies.

Despite the larger gas mass suppression in the smallest halos, the results of the {\it Jeans mass} model are qualitatively the same as those from the \textit{Photoionization} model. However, (low-mass) galaxies of a given halo mass have lower UV luminosities in the {\it Early heating} model.

\section{Conclusions and Discussion}
This work aims at quantifying the impact of cosmic variance on the observed UV Luminosity Function (UV LF), the Halo Mass Function (HMF), and the Stellar Mass Function (SMF) at $z \sim 6-12$ using the results of the \textlcsc{Astraeus} (semi-numerical rAdiative tranSfer coupling of galaxy formaTion andReionization in N-body dArk mattEr simUlationS) framework. This framework couples a state-of-the-art N-body simulation ($160 h^{-1}$ Mpc with a mass resolution of $6.2 \times 10^6 h^{-1}$ $\Msun$) with the \textlcsc{Delphi} semi-analytic model of galaxy formation and the \textlcsc{CIFOG} (Code to compute ionization field from density fields and source catalogue) semi-numerical IGM reionization scheme.

We studied three forthcoming surveys (JADES-deep, JADES-medium and WFIRST1) and different reionization scenarios. We find that for the JADES-deep survey, the cosmic variance ($\zeta$) increases from about 10\% to 100\% as $M_h$ increases from $10^9$ $\Msun$ to $10^{11}$ $\Msun$. We find that the contribution from reionization modelled with our different scenarios play a minor role in the cosmic variance. Most of the cosmic variance is indeed completely driven by the underlying density field. While $\zeta \lsim 20\%$ for $\muv \sim -12$ to $-15$ LBGs at $z \sim 6-12$ in all the UV feedback models studied, it increases above $100\%$ for $\muv \sim -17.5 ~ (-22)$ at $z =12~ (8)$. As expected, the cosmic variance decreases with an increasing survey area: an increase in survey area by a factor 4 from the JADES-deep to the JADES medium survey results in a decrease in the cosmic variance by a factor $\sim$ 2, with this scaling being roughly independent of redshift. Furthermore, our analysis suggests that to minimize the cosmic variance ($\lsim$ 10\%), the survey area should be at least $\sim$ 1000 arcmin$^2$.

We find that the faint end slope ($\alpha$) of the UV LF becomes increasingly shallower with decreasing redshift for all the reionization models explored. The redshift evolution of $\alpha $ is the shallowest for the \textit{Jeans mass} model, where the gas mass of low-mass halos is instantaneously suppressed by reionization feedback, as compared to the \textit{Photoionization} and \textit{Early heating} models. Although the values of $\alpha$ are comparable for all three models at $z<9$, at $z = 9-12$, even accounting for cosmic variance, the value of $\alpha$ in the \textit{Jeans mass} model is distinguishable from those in the other models considered in this work. Therefore, integrating down to $\muv = -15$, a reasonable limit for the JWST, would be sufficient to differentiate between these different UV feedback models.

We also explored the environments of LBGs in the EoR. As expected, we found the most luminous LBGs to live in the most ionized and over-dense regions. The ionization fields for the \textit{Early Heating} model differ strongly from the other two models especially in the initial stages of reionization ($z >10$) as a result of low-mass galaxies providing a larger fraction of ionizing photons. This naturally results in a larger bulk of low-luminosity galaxies lying in ionised regions, compared to the other two models. Finally, this model shows an upturn in the $1+\delta-\chi_{HI}$ relation at $1+\delta>3$ (compared to the downturn in the other two models), which is driven by the lower escape fraction of \HI ionizing photons from the most massive galaxies.

Finally, we also provide a public software tool to compute the UV LF, HMF, and SMF cosmic variance for different redshifts, redshift intervals and survey areas\footnote{https://github.com/grazianoucci/cosmic\_variance}.

\section*{Acknowledgments}
The authors thank the anonymous referees for their comments that improved the quality of the paper. GU, PD, AH, GY, SG and LL acknowledge support from the European Research Council's starting grant ERC StG-717001 (\quotes{DELPHI}). PD also acknowledges support from the NWO's VIDI grant (``ODIN"; 016.vidi.189.162) and the European Commission's and University of Groningen's CO-FUND Rosalind Franklin program. GY acknowledges financial support from MINECO/FEDER under project grant AYA2015-63810-P and MICIU/FEDER under project grant  PGC2018-094975-C21. The authors wish to thank V. Springel for allowing us to use the L-Gadget2 code to run the different Multidark simulation boxes, including the VSMDPL used in this work. The VSMDPL and ESMDPL simulations have been performed at LRZ Munich within the project pr87yi. The CosmoSim database (www.cosmosim.org) provides access to the simulation and the Rockstar data. The database is a service by the Leibniz Institute for Astrophysics Potsdam (AIP).

\bibliographystyle{mn2e}
\bibliography{delphi}

\begin{thebibliography}{73}
\expandafter\ifx\csname natexlab\endcsname\relax\def\natexlab#1{#1}\fi

\bibitem[{{Atek} {et~al}\mbox{.}(2014){Atek}, {Richard}, {Kneib}, {Clement},
  {Egami}, {Ebeling}, {Jauzac}, {Jullo}, {Laporte}, {Limousin}, \&
  {Natarajan}}]{atek2014}
{Atek} H. {et~al.}, 2014, \apj, 786, 60

\bibitem[{{Atek} {et~al}\mbox{.}(2018){Atek}, {Richard}, {Kneib}, \&
  {Schaerer}}]{atek2018}
{Atek} H., {Richard} J., {Kneib} J.-P., {Schaerer} D., 2018, \mnras, 479, 5184

\bibitem[{{Barkana} \& {Loeb}(1999)}]{barkana1999}
{Barkana} R., {Loeb} A., 1999, \apj, 523, 54

\bibitem[{{Behroozi} {et~al}\mbox{.}(2013{\natexlab{a}}){Behroozi}, {Wechsler},
  \& {Wu}}]{behroozi2013}
{Behroozi} P.~S., {Wechsler} R.~H., {Wu} H.-Y., 2013{\natexlab{a}}, \apj, 762,
  109

\bibitem[{{Behroozi} {et~al}\mbox{.}(2013{\natexlab{b}}){Behroozi}, {Wechsler},
  {Wu}, {Busha}, {Klypin}, \& {Primack}}]{behroozi2013_trees}
{Behroozi} P.~S., {Wechsler} R.~H., {Wu} H.-Y., {Busha} M.~T., {Klypin} A.~A.,
  {Primack} J.~R., 2013{\natexlab{b}}, \apj, 763, 18

\bibitem[{{Bhowmick} {et~al}\mbox{.}(2019){Bhowmick}, {Somerville}, {DiMatteo},
  {Wilkins}, {Feng}, \& {Tenneti}}]{Bhowmick2019}
{Bhowmick} A.~K., {Somerville} R.~S., {DiMatteo} T., {Wilkins} S., {Feng} Y.,
  {Tenneti} A., 2019, arXiv e-prints, arXiv:1908.02787

\bibitem[{{Bouwens} {et~al}\mbox{.}(2012){Bouwens}, {Bradley}, {Zitrin}, {Coe},
  {Franx}, {Zheng}, {Smit}, {Host}, {Postman}, {Moustakas}, {Labbe},
  {Carrasco}, {Molino}, {Donahue}, {Kelson}, {Meneghetti}, {Jha}, {Benitez},
  {Lemze}, {Umetsu}, {Broadhurst}, {Moustakas}, {Rosati}, {Bartelmann}, {Ford},
  {Graves}, {Grillo}, {Infante}, {Jiminez-Teja}, {Jouvel}, {Lahav}, {Maoz},
  {Medezinski}, {Melchior}, {Merten}, {Nonino}, {Ogaz}, \&
  {Seitz}}]{bouwens2012}
{Bouwens} R. {et~al.}, 2012, ArXiv e-prints

\bibitem[{{Bouwens} {et~al}\mbox{.}(2015){Bouwens}, {Illingworth}, {Oesch},
  {Trenti}, {Labb{\'e}}, {Bradley}, {Carollo}, {van Dokkum}, {Gonzalez},
  {Holwerda}, {Franx}, {Spitler}, {Smit}, \& {Magee}}]{bouwens2015b}
{Bouwens} R.~J. {et~al.}, 2015, \apj, 803, 34

\bibitem[{{Bouwens} {et~al}\mbox{.}(2017){Bouwens}, {Oesch}, {Illingworth},
  {Ellis}, \& {Stefanon}}]{bouwens2017}
{Bouwens} R.~J., {Oesch} P.~A., {Illingworth} G.~D., {Ellis} R.~S., {Stefanon}
  M., 2017, \apj, 843, 129

\bibitem[{{Bouwens} {et~al}\mbox{.}(2016){Bouwens}, {Oesch}, {Labb{\'e}},
  {Illingworth}, {Fazio}, {Coe}, {Holwerda}, {Smit}, {Stefanon}, {van Dokkum},
  {Trenti}, {Ashby}, {Huang}, {Spitler}, {Straatman}, {Bradley}, \&
  {Magee}}]{bouwens2016b}
{Bouwens} R.~J. {et~al.}, 2016, \apj, 830, 67

\bibitem[{{Bowler} {et~al}\mbox{.}(2015){Bowler}, {Dunlop}, {McLure},
  {McCracken}, {Milvang-Jensen}, {Furusawa}, {Taniguchi}, {Le F{\`e}vre},
  {Fynbo}, {Jarvis}, \& {H{\"a}u{\ss}ler}}]{bowler2015}
{Bowler} R.~A.~A. {et~al.}, 2015, \mnras, 452, 1817

\bibitem[{{Bradley} {et~al}\mbox{.}(2012){Bradley}, {Trenti}, {Oesch},
  {Stiavelli}, {Treu}, {Bouwens}, {Shull}, {Holwerda}, \&
  {Pirzkal}}]{bradley2012}
{Bradley} L.~D. {et~al.}, 2012, \apj, 760, 108

\bibitem[{{Castellano} {et~al}\mbox{.}(2016){Castellano}, {Yue}, {Ferrara},
  {Merlin}, {Fontana}, {Amor{\'\i}n}, {Grazian}, {M{\'a}rmol-Queralto},
  {Micha{\l}owski}, {Mortlock}, {Paris}, {Parsa}, {Pilo}, \&
  {Santini}}]{castellano2016}
{Castellano} M. {et~al.}, 2016, \apjl, 823, L40

\bibitem[{{Chardin} {et~al}\mbox{.}(2017){Chardin}, {Puchwein}, \&
  {Haehnelt}}]{chardin2017}
{Chardin} J., {Puchwein} E., {Haehnelt} M.~G., 2017, \mnras, 465, 3429

\bibitem[{{Choudhury} \& {Dayal}(2019)}]{choudhury2019}
{Choudhury} T.~R., {Dayal} P., 2019, \mnras, 482, L19

\bibitem[{{Choudhury} \& {Ferrara}(2007)}]{choudhury2007}
{Choudhury} T.~R., {Ferrara} A., 2007, \mnras, 380, L6

\bibitem[{{Dawoodbhoy} {et~al}\mbox{.}(2018){Dawoodbhoy}, {Shapiro}, {Ocvirk},
  {Aubert}, {Gillet}, {Choi}, {Iliev}, {Teyssier}, {Yepes}, {Gottl{\"o}ber},
  {D'Aloisio}, {Park}, \& {Hoffman}}]{dawoodbhoy2018}
{Dawoodbhoy} T. {et~al.}, 2018, \mnras, 480, 1740

\bibitem[{{Dayal} {et~al}\mbox{.}(2017{\natexlab{a}}){Dayal}, {Choudhury},
  {Bromm}, \& {Pacucci}}]{dayal2017a}
{Dayal} P., {Choudhury} T.~R., {Bromm} V., {Pacucci} F., 2017{\natexlab{a}},
  \apj, 836, 16

\bibitem[{{Dayal} {et~al}\mbox{.}(2017{\natexlab{b}}){Dayal}, {Choudhury},
  {Pacucci}, \& {Bromm}}]{dayal2017b}
{Dayal} P., {Choudhury} T.~R., {Pacucci} F., {Bromm} V., 2017{\natexlab{b}},
  \mnras, 472, 4414

\bibitem[{{Dayal} \& {Ferrara}(2018)}]{dayal2018}
{Dayal} P., {Ferrara} A., 2018, \physrep, 780, 1

\bibitem[{{Dayal} {et~al}\mbox{.}(2014){Dayal}, {Ferrara}, {Dunlop}, \&
  {Pacucci}}]{dayal2014}
{Dayal} P., {Ferrara} A., {Dunlop} J.~S., {Pacucci} F., 2014, \mnras, 445, 2545

\bibitem[{{Dayal} {et~al}\mbox{.}(2015){Dayal}, {Mesinger}, \&
  {Pacucci}}]{dayal2015}
{Dayal} P., {Mesinger} A., {Pacucci} F., 2015, \apj, 806, 67

\bibitem[{{Dayal} {et~al}\mbox{.}(2020){Dayal}, {Volonteri}, {Choudhury},
  {Schneider}, {Trebitsch}, {Gnedin}, {Atek}, {Hirschmann}, \&
  {Reines}}]{dayal2020}
{Dayal} P. {et~al.}, 2020, arXiv e-prints, arXiv:2001.06021

\bibitem[{{Driver} \& {Robotham}(2010)}]{Driver2010}
{Driver} S.~P., {Robotham} A. S.~G., 2010, \mnras, 407, 2131

\bibitem[{{Duncan} \& {Conselice}(2015)}]{duncan2015}
{Duncan} K., {Conselice} C.~J., 2015, \mnras, 451, 2030

\bibitem[{{Duncan} {et~al}\mbox{.}(2014){Duncan}, {Conselice}, {Mortlock},
  {Hartley}, {Guo}, {Ferguson}, {Dav{\'e}}, {Lu}, {Ownsworth}, {Ashby},
  {Dekel}, {Dickinson}, {Faber}, {Giavalisco}, {Grogin}, {Kocevski},
  {Koekemoer}, {Somerville}, \& {White}}]{duncan2014}
{Duncan} K. {et~al.}, 2014, \mnras, 444, 2960

\bibitem[{{Eldridge} {et~al}\mbox{.}(2017){Eldridge}, {Stanway}, {Xiao},
  {McClelland }, {Taylor}, {Ng}, {Greis}, \& {Bray}}]{eldridge2017}
{Eldridge} J.~J., {Stanway} E.~R., {Xiao} L., {McClelland } L.~A.~S., {Taylor}
  G., {Ng} M., {Greis} S.~M.~L., {Bray} J.~C., 2017, \pasa, 34, e058

\bibitem[{{Finkelstein} {et~al}\mbox{.}(2019){Finkelstein}, {D'Aloisio},
  {Paardekooper}, {Ryan}, {Behroozi}, {Finlator}, {Livermore}, {Upton
  Sanderbeck}, {Dalla Vecchia}, \& {Khochfar}}]{finkelstein2019}
{Finkelstein} S.~L. {et~al.}, 2019, \apj, 879, 36

\bibitem[{{Finkelstein} {et~al}\mbox{.}(2015{\natexlab{a}}){Finkelstein},
  {Dunlop}, {Le Fevre}, \& {Wilkins}}]{finkelstein2015arXiv}
{Finkelstein} S.~L., {Dunlop} J., {Le Fevre} O., {Wilkins} S.,
  2015{\natexlab{a}}, arXiv e-prints, arXiv:1512.04530

\bibitem[{{Finkelstein} {et~al}\mbox{.}(2012){Finkelstein}, {Papovich},
  {Salmon}, {Finlator}, {Dickinson}, {Ferguson}, {Giavalisco}, {Koekemoer},
  {Reddy}, {Bassett}, {Conselice}, {Dunlop}, {Faber}, {Grogin}, {Hathi},
  {Kocevski}, {Lai}, {Lee}, {McLure}, {Mobasher}, \&
  {Newman}}]{finkelstein2012}
{Finkelstein} S.~L. {et~al.}, 2012, \apj, 756, 164

\bibitem[{{Finkelstein} {et~al}\mbox{.}(2015{\natexlab{b}}){Finkelstein},
  {Ryan}, {Papovich}, {Dickinson}, {Song}, {Somerville}, {Ferguson}, {Salmon},
  {Giavalisco}, {Koekemoer}, {Ashby}, {Behroozi}, {Castellano}, {Dunlop},
  {Faber}, {Fazio}, {Fontana}, {Grogin}, {Hathi}, {Jaacks}, {Kocevski},
  {Livermore}, {McLure}, {Merlin}, {Mobasher}, {Newman}, {Rafelski}, {Tilvi},
  \& {Willner}}]{finkelstein2015}
{Finkelstein} S.~L. {et~al.}, 2015{\natexlab{b}}, \apj, 810, 71

\bibitem[{{Finlator} {et~al}\mbox{.}(2011){Finlator}, {Dav{\'e}}, \&
  {{\"O}zel}}]{finlator2011}
{Finlator} K., {Dav{\'e}} R., {{\"O}zel} F., 2011, \apj, 743, 169

\bibitem[{{Giallongo} {et~al}\mbox{.}(2015){Giallongo}, {Grazian}, {Fiore},
  {Fontana}, {Pentericci}, {Vanzella}, {Dickinson}, {Kocevski}, {Castellano},
  {Cristiani}, {Ferguson}, {Finkelstein}, {Grogin}, {Hathi}, {Koekemoer},
  {Newman}, \& {Salvato}}]{giallongo2015}
{Giallongo} E. {et~al.}, 2015, \aap, 578, A83

\bibitem[{{Gnedin}(2000)}]{gnedin2000}
{Gnedin} N.~Y., 2000, \apj, 542, 535

\bibitem[{{Gonz{\'a}lez} {et~al}\mbox{.}(2011){Gonz{\'a}lez}, {Labb{\'e}},
  {Bouwens}, {Illingworth}, {Franx}, \& {Kriek}}]{gonzalez2011}
{Gonz{\'a}lez} V., {Labb{\'e}} I., {Bouwens} R.~J., {Illingworth} G., {Franx}
  M., {Kriek} M., 2011, \apjl, 735, L34

\bibitem[{{Hasegawa} \& {Semelin}(2013)}]{hasegawa2013}
{Hasegawa} K., {Semelin} B., 2013, \mnras, 428, 154

\bibitem[{{Hutter}(2018)}]{hutter2018}
{Hutter} A., 2018, \mnras, 477, 1549

\bibitem[{{Hutter} {et~al}\mbox{.}(2017){Hutter}, {Dayal}, {M{\"u}ller}, \&
  {Trott}}]{hutter2017}
{Hutter} A., {Dayal} P., {M{\"u}ller} V., {Trott} C.~M., 2017, \apj, 836, 176

\bibitem[{{Hutter} {et~al}\mbox{.}(2021){Hutter}, {Dayal}, {Yepes},
  {Gottl{\"o}ber}, {Legrand}, \& {Ucci}}]{astraeus1}
{Hutter} A., {Dayal} P., {Yepes} G., {Gottl{\"o}ber} S., {Legrand} L., {Ucci}
  G., 2021, \mnras, 503, 3698

\bibitem[{{Ishigaki} {et~al}\mbox{.}(2018){Ishigaki}, {Kawamata}, {Ouchi},
  {Oguri}, {Shimasaku}, \& {Ono}}]{ishigaki2018}
{Ishigaki} M., {Kawamata} R., {Ouchi} M., {Oguri} M., {Shimasaku} K., {Ono} Y.,
  2018, \apj, 854, 73

\bibitem[{{Kemp} {et~al}\mbox{.}(2019){Kemp}, {Dunlop}, {McLure}, {Schreiber},
  {Carnall}, \& {Cullen}}]{kemp2019}
{Kemp} T.~W., {Dunlop} J.~S., {McLure} R.~J., {Schreiber} C., {Carnall} A.~C.,
  {Cullen} F., 2019, \mnras, 486, 3087

\bibitem[{{Leitherer} {et~al}\mbox{.}(1999){Leitherer}, {Schaerer}, {Goldader},
  {Gonz{\'a}lez Delgado}, {Robert}, {Kune}, {de Mello}, {Devost}, \&
  {Heckman}}]{leitherer1999}
{Leitherer} C. {et~al.}, 1999, \apjs, 123, 3

\bibitem[{{Liu} {et~al}\mbox{.}(2016){Liu}, {Mutch}, {Angel}, {Duffy}, {Geil},
  {Poole}, {Mesinger}, \& {Wyithe}}]{liu2016}
{Liu} C., {Mutch} S.~J., {Angel} P.~W., {Duffy} A.~R., {Geil} P.~M., {Poole}
  G.~B., {Mesinger} A., {Wyithe} J.~S.~B., 2016, \mnras, 462, 235

\bibitem[{{Livermore} {et~al}\mbox{.}(2017){Livermore}, {Finkelstein}, \&
  {Lotz}}]{livermore2017}
{Livermore} R.~C., {Finkelstein} S.~L., {Lotz} J.~M., 2017, \apj, 835, 113

\bibitem[{{Mac Low} \& {Ferrara}(1999)}]{maclow-ferrara1999}
{Mac Low} M.-M., {Ferrara} A., 1999, \apj, 513, 142

\bibitem[{{Madau} \& {Haardt}(2015)}]{madau2015}
{Madau} P., {Haardt} F., 2015, \apjl, 813, L8

\bibitem[{{McLure} {et~al}\mbox{.}(2013){McLure}, {Dunlop}, {Bowler},
  {Curtis-Lake}, {Schenker}, {Ellis}, {Robertson}, {Koekemoer}, {Rogers},
  {Ono}, {Ouchi}, {Charlot}, {Wild}, {Stark}, {Furlanetto}, {Cirasuolo}, \&
  {Targett}}]{mclure2013}
{McLure} R.~J. {et~al.}, 2013, \mnras, 432, 2696

\bibitem[{{Mitra} {et~al}\mbox{.}(2018){Mitra}, {Choudhury}, \&
  {Ferrara}}]{mitra2018}
{Mitra} S., {Choudhury} T.~R., {Ferrara} A., 2018, \mnras, 473, 1416

\bibitem[{{Moster} {et~al}\mbox{.}(2011){Moster}, {Somerville}, {Newman}, \&
  {Rix}}]{Moster2011}
{Moster} B.~P., {Somerville} R.~S., {Newman} J.~A., {Rix} H.-W., 2011, \apj,
  731, 113

\bibitem[{{Mutch} {et~al}\mbox{.}(2016){Mutch}, {Geil}, {Poole}, {Angel},
  {Duffy}, {Mesinger}, \& {Wyithe}}]{mutch2016}
{Mutch} S.~J., {Geil} P.~M., {Poole} G.~B., {Angel} P.~W., {Duffy} A.~R.,
  {Mesinger} A., {Wyithe} J. S.~B., 2016, \mnras, 462, 250

\bibitem[{{Naoz} {et~al}\mbox{.}(2013){Naoz}, {Yoshida}, \&
  {Gnedin}}]{naoz2013}
{Naoz} S., {Yoshida} N., {Gnedin} N.~Y., 2013, \apj, 763, 27

\bibitem[{{Oesch} {et~al}\mbox{.}(2013){Oesch}, {Bouwens}, {Illingworth},
  {Labb{\'e}}, {Franx}, {van Dokkum}, {Trenti}, {Stiavelli}, {Gonzalez}, \&
  {Magee}}]{oesch2013}
{Oesch} P.~A. {et~al.}, 2013, \apj, 773, 75

\bibitem[{{Oesch} {et~al}\mbox{.}(2018){Oesch}, {Bouwens}, {Illingworth},
  {Labb{\'e}}, \& {Stefanon}}]{oesch2018}
{Oesch} P.~A., {Bouwens} R.~J., {Illingworth} G.~D., {Labb{\'e}} I., {Stefanon}
  M., 2018, \apj, 855, 105

\bibitem[{{Petkova} \& {Springel}(2011)}]{petkova2011}
{Petkova} M., {Springel} V., 2011, \mnras, 412, 935

\bibitem[{{Planck Collaboration} {et~al}\mbox{.}(2016){Planck Collaboration},
  {Adam}, {Ade}, {Aghanim}, {Akrami}, {Alves}, {Arg{\"u}eso}, {Arnaud},
  {Arroja}, {Ashdown}, \& et~al.}]{planck2016}
{Planck Collaboration} {et~al.}, 2016, \aap, 594, A1

\bibitem[{{Qin} {et~al}\mbox{.}(2017){Qin}, {Mutch}, {Poole}, {Liu}, {Duffy},
  {Geil}, {Angel}, {Mesinger}, \& {Wyithe}}]{qin2017}
{Qin} Y. {et~al.}, 2017, ArXiv e-prints

\bibitem[{{Rafelski} {et~al}\mbox{.}(2012){Rafelski}, {Wolfe}, {Prochaska},
  {Neeleman}, \& {Mendez}}]{rafelski2012}
{Rafelski} M., {Wolfe} A.~M., {Prochaska} J.~X., {Neeleman} M., {Mendez} A.~J.,
  2012, \apj, 755, 89

\bibitem[{{Razoumov} \& {Sommer-Larsen}(2010)}]{razoumov2010}
{Razoumov} A.~O., {Sommer-Larsen} J., 2010, \apj, 710, 1239

\bibitem[{{Rieke} {et~al}\mbox{.}(2019){Rieke}, {Arribas}, {Bunker}, {Charlot},
  {Finkelstein}, {Maiolino}, {Robertson}, {Willott}, {Windhorst}, {Eisenstein},
  {Nelson}, {Tacchella}, {Egami}, {Endsley}, {Frye}, {Hainline}, {Hviding},
  {Rieke}, {Williams}, {Willmer}, \& {Woodrum}}]{rieke2019}
{Rieke} M. {et~al.}, 2019, \baas, 51, 45

\bibitem[{{Robertson} {et~al}\mbox{.}(2015){Robertson}, {Ellis}, {Furlanetto},
  \& {Dunlop}}]{robertson2015}
{Robertson} B.~E., {Ellis} R.~S., {Furlanetto} S.~R., {Dunlop} J.~S., 2015,
  \apjl, 802, L19

\bibitem[{{Salpeter}(1955)}]{salpeter1955}
{Salpeter} E.~E., 1955, \apj, 121, 161

\bibitem[{{Salvaterra} {et~al}\mbox{.}(2011){Salvaterra}, {Ferrara}, \&
  {Dayal}}]{salvaterra2011}
{Salvaterra} R., {Ferrara} A., {Dayal} P., 2011, \mnras, 414, 847

\bibitem[{{Schenker} {et~al}\mbox{.}(2013){Schenker}, {Robertson}, {Ellis},
  {Ono}, {McLure}, {Dunlop}, {Koekemoer}, {Bowler}, {Ouchi}, {Curtis-Lake},
  {Rogers}, {Schneider}, {Charlot}, {Stark}, {Furlanetto}, \&
  {Cirasuolo}}]{schenker2013}
{Schenker} M.~A. {et~al.}, 2013, \apj, 768, 196

\bibitem[{{Seiler} {et~al}\mbox{.}(2018){Seiler}, {Hutter}, {Sinha}, \&
  {Croton}}]{seiler2018}
{Seiler} J., {Hutter} A., {Sinha} M., {Croton} D., 2018, \mnras, 480, L33

\bibitem[{{Shapiro} {et~al}\mbox{.}(2004){Shapiro}, {Iliev}, \&
  {Raga}}]{shapiro2004}
{Shapiro} P.~R., {Iliev} I.~T., {Raga} A.~C., 2004, \mnras, 348, 753

\bibitem[{{Sobacchi} \& {Mesinger}(2013)}]{sobacchi2013a}
{Sobacchi} E., {Mesinger} A., 2013, \mnras, 432, L51

\bibitem[{{Song} {et~al}\mbox{.}(2016){Song}, {Finkelstein}, {Ashby},
  {Grazian}, {Lu}, {Papovich}, {Salmon}, {Somerville}, {Dickinson}, {Duncan},
  {Faber}, {Fazio}, {Ferguson}, {Fontana}, {Guo}, {Hathi}, {Lee}, {Merlin}, \&
  {Willner}}]{song2016}
{Song} M. {et~al.}, 2016, \apj, 825, 5

\bibitem[{{Stark}(2016)}]{stark2016}
{Stark} D.~P., 2016, \araa, 54, 761

\bibitem[{{Trenti} \& {Stiavelli}(2008)}]{Trenti2008}
{Trenti} M., {Stiavelli} M., 2008, \apj, 676, 767

\bibitem[{{Volonteri} \& {Gnedin}(2009)}]{volonteri2009}
{Volonteri} M., {Gnedin} N.~Y., 2009, \apj, 703, 2113

\bibitem[{{Williams} {et~al}\mbox{.}(2018){Williams}, {Curtis-Lake},
  {Hainline}, {Chevallard}, {Robertson}, {Charlot}, {Endsley}, {Stark},
  {Willmer}, {Alberts}, {Amorin}, {Arribas}, {Baum}, {Bunker}, {Carniani},
  {Crand all}, {Egami}, {Eisenstein}, {Ferruit}, {Husemann}, {Maseda},
  {Maiolino}, {Rawle}, {Rieke}, {Smit}, {Tacchella}, \&
  {Willott}}]{williams2018}
{Williams} C.~C. {et~al.}, 2018, \apjs, 236, 33

\bibitem[{{Willott} {et~al}\mbox{.}(2013){Willott}, {McLure}, {Hibon},
  {Bielby}, {McCracken}, {Kneib}, {Ilbert}, {Bonfield}, {Bruce}, \&
  {Jarvis}}]{willott2013}
{Willott} C.~J. {et~al.}, 2013, \aj, 145, 4

\bibitem[{{Yue} {et~al}\mbox{.}(2018){Yue}, {Castellano}, {Ferrara}, {Fontana},
  {Merlin}, {Amor{\'\i}n}, {Grazian}, {M{\'a}rmol-Queralto}, {Micha{\l}owski},
  {Mortlock}, {Paris}, {Parsa}, {Pilo}, {Santini}, \& {Di
  Criscienzo}}]{yue2018}
{Yue} B. {et~al.}, 2018, \apj, 868, 115

\end{thebibliography}

\appendix
\section{Convergence test}
In this appendix, we study the  effects of limited box sizes on our estimates of cosmic variances. To this end we use a smaller box simulated with higher resolution as well as sub-boxes of the original simulation with the same resolution. In order to estimate at which halo masses, stellar masses and UV luminosities our results are robust, we compare the cosmic variances for JADES-DEEP surveys derived from the \textlcsc{VSMDPL} ($160h^{-1}$ cMpc) and the Extemely Small Multidark Planck (\textlcsc{ESMDPL}, $64h^{-1}$ cMpc) simulations. Using the models analysed in Appendix B of \cite{astraeus1} (instantaneous SN feedback and no radiative feedback), we repeat the analysis described in Section \ref{sec_surveys} for the \textlcsc{VSMDPL} and \textlcsc{ESMDPL} simulations. While we derive the cosmic variance by considering JADES-DEEP volumes within the entire \textlcsc{ESMDPL} simulation box, we derive 8 cosmic variance estimates from 8 non-overlapping \textlcsc{ESMDPL}-sized sub-boxes within the \textlcsc{VSMDPL} simulation box.

\begin{figure}
	\centering
	\includegraphics[width=1.0\linewidth]{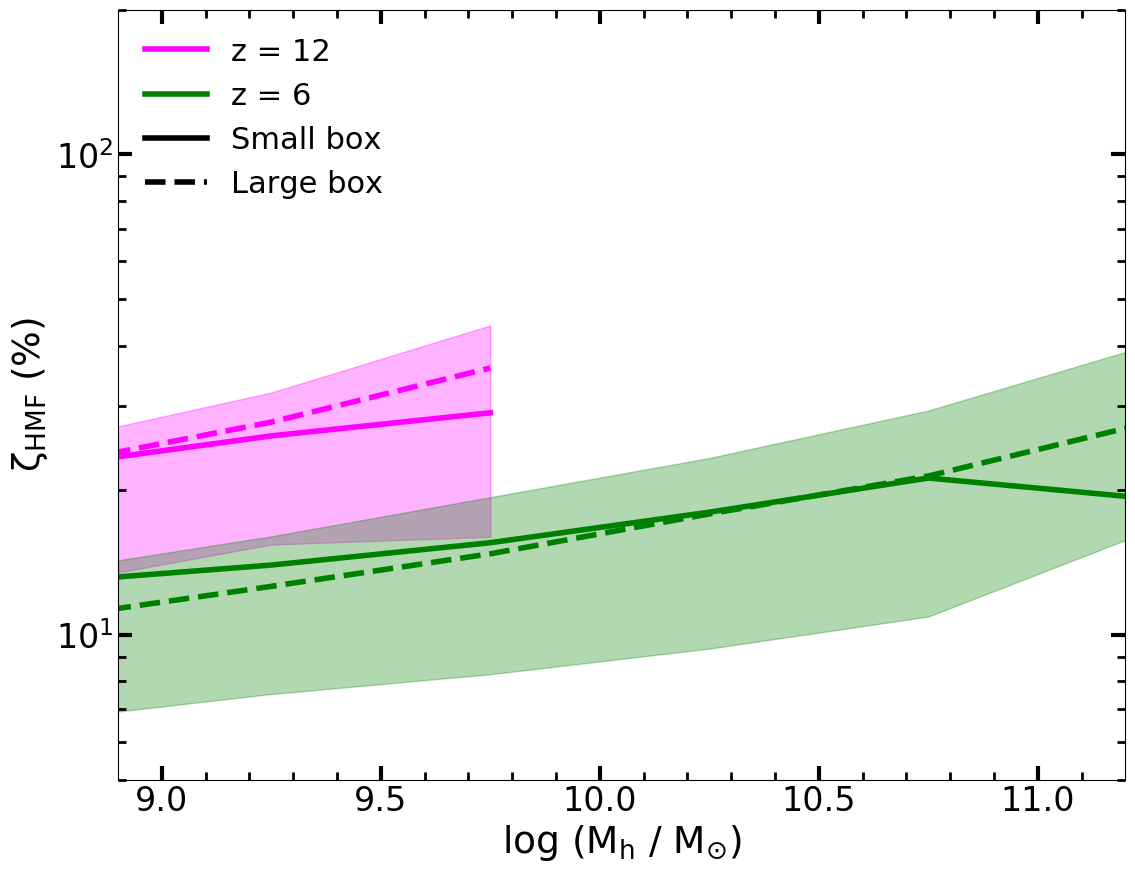}
	\caption{Comparison between the HMF cosmic variance obtained for the JADES-DEEP survey using data from the large box simulation (dashed line; \textlcsc{VSMDPL}: 160$h^{-1}$ cMpc) and small box simulation (solid line; \textlcsc{ESMDPL}, 64$h^{-1}$ cMpc) at $z \sim 6-12$. The shaded area denotes the cosmic variance values spanned by 8 sub-boxes inside \textlcsc{VSMDPL} each of which have a box length equal to 64$h^{-1}$ cMpc.}
	\label{fig:hmf_conv}
\end{figure}

\begin{figure}
	\centering
	\includegraphics[width=1.0\linewidth]{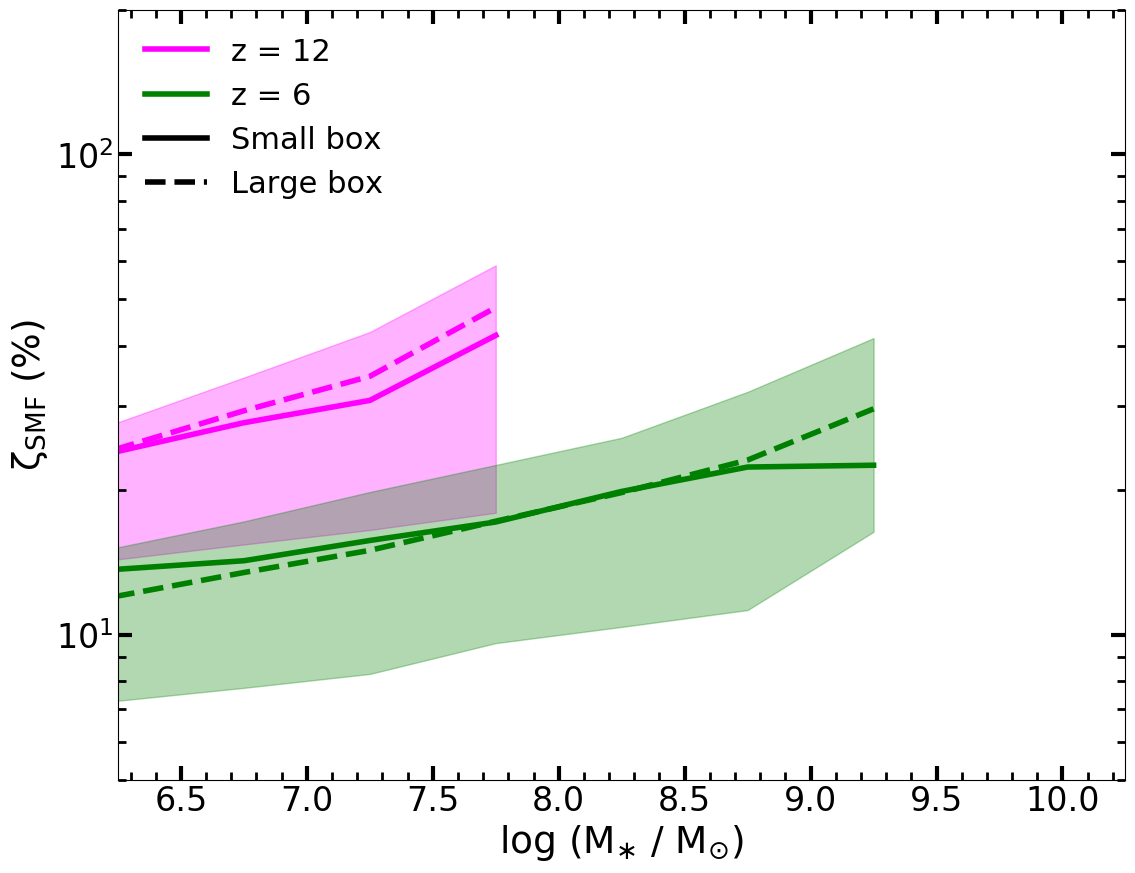}
	\caption{Comparison between the SMF cosmic variance obtained for the JADES-DEEP survey using data from the large box simulation (dashed line; \textlcsc{VSMDPL}, 160$h^{-1}$ cMpc) and small box simulation (solid line; \textlcsc{ESMDPL}, 64$h^{-1}$ cMpc) at $z \sim 6-12$. The shaded area denotes the cosmic variance values spanned by 8 sub-boxes inside \textlcsc{VSMDPL} each of which have a box length equal to 64$h^{-1}$ cMpc.}
	\label{fig:smf_conv}
\end{figure}

\begin{figure}
	\centering
	\includegraphics[width=1.0\linewidth]{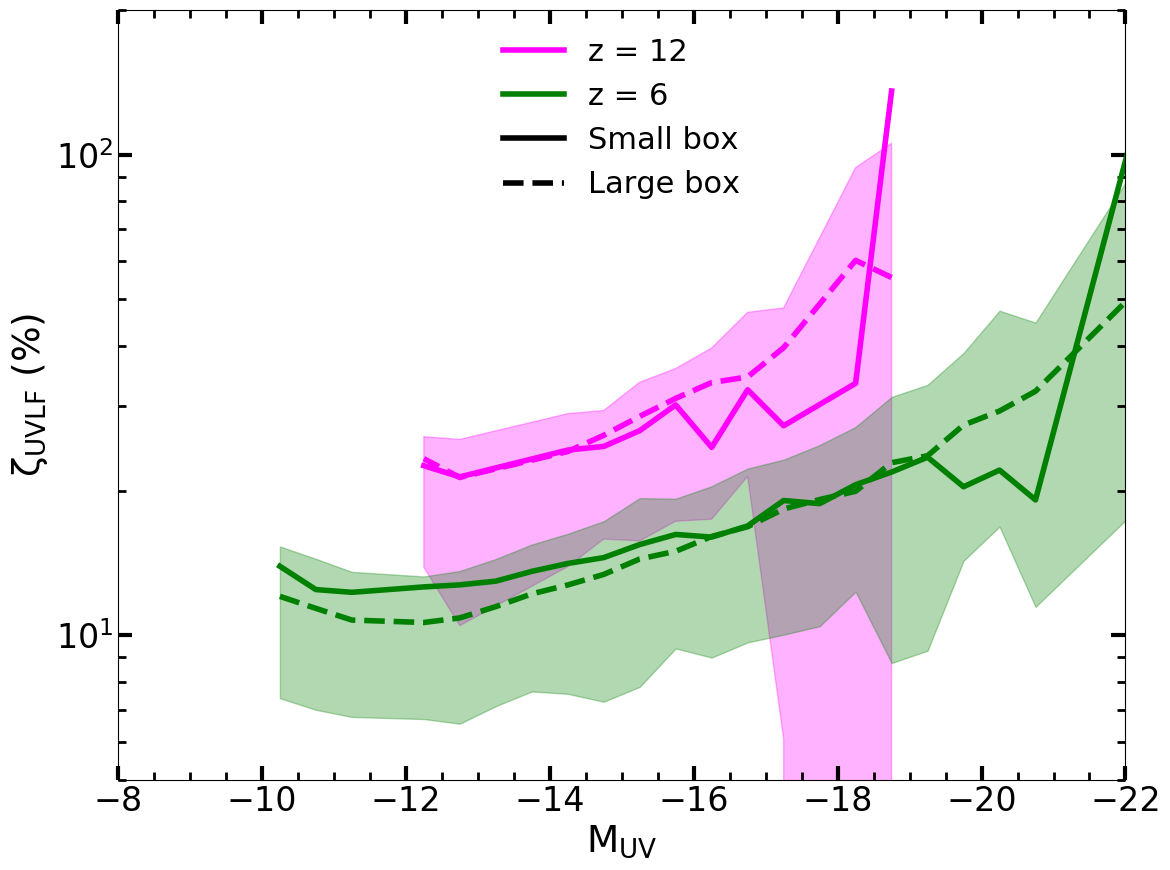}
	\caption{Comparison between the UVLF cosmic variance obtained for the JADES-DEEP survey using data from the large box simulation (dashed line; \textlcsc{VSMDPL}, 160$h^{-1}$ cMpc) and small box simulation (solid line; \textlcsc{ESMDPL}, 64$h^{-1}$ cMpc) at $z \sim 6-12$. The shaded area denotes the cosmic variance values spanned by 8 sub-boxes inside \textlcsc{VSMDPL} each of which have a box length equal to 64$h^{-1}$ cMpc.}
	\label{fig:uvlf_conv}
\end{figure}

In Figs. \ref{fig:hmf_conv}, \ref{fig:smf_conv} and \ref{fig:uvlf_conv}, for clarity, we show the HMF, SMF and UV LF cosmic variance at the minimum and maximum redshifts studied i.e. $z \sim 6$ and $12$, respectively. Here, we have limited the underlying galaxy sample to (1) galaxies whose properties have converged ($M_h \geq 10^{8.6} \msun$) and (2) halo masses for which the HMFs of the \textlcsc{VSMDPL} and \textlcsc{ESMDPL} simulations agree: $M_h \sim 10^{8.75-11.25} \msun (10^{8.75-9.75}$) at $z \simeq 6$ (12). We see that the cosmic variance range spanned by the sub-volumes of the \textlcsc{VSMDPL} simulations (shaded coloured regions) encompass the cosmic variance derived from the \textlcsc{ESMDPL} simulation box (solid lines) and results from the whole \textlcsc{VSMDPL box} (dashed lines). For example, the variance between the different sub-volumes of the \textlcsc{VSMDPL} spans a factor of $\sim$ 2.2 for the HMFs, SMFs and UV LFs, irrespective of redshift. On the other hand, the variance between the full \textlcsc{ESMDPL} and \textlcsc{VSMDPL} boxes (solid and dashed lines in the figures) only differ by at most a factor of $\sim$ 0.8-1.5 at $z \sim 6-12$. Finally, the \textlcsc{VSMDPL} sub-volumes can be used to obtain an indication of the uncertainty in the variance expected if the JADES-deep surveys were to be constructed from these small-volumes. For example, the variance in the halo mass function increases by a factor of $\sim$ 1.3 from $\sim 24$\% to $\sim$ 32\% as the halo mass increases from $M_h \sim 10^9\msun$ to $10^{9.5}\msun$ at $z \sim 12$ and from $\sim$ 12\% to $\sim$ 20\% as the halo mass increases from $M_h \sim 10^9 \msun$ to $10^{10.5}\msun$ at $z \sim 6$. This, coupled with the baryonic assembly histories of galaxies, leads to a variance that ranges between 15-34\% for $\muv \sim -15$ at $z \sim 12$, and between 8-15\% (15-40\%) at $\muv \sim -14 \, (-20)$ at $z \sim 6$.

\label{lastpage}
\end{document}